\documentclass[twocolumn,superscriptaddress,english,prl,showpacs,longbibliography]{revtex4-2}

\usepackage[colorlinks=true,urlcolor=blue,citecolor=blue,linkcolor=blue]{hyperref} 
\usepackage{graphicx}
\usepackage{dcolumn}
\usepackage{adjustbox}
\usepackage{bm}
\usepackage{color}
\usepackage{physics}
\usepackage{float}
\makeatletter
\let\newfloat\newfloat@ltx
\makeatother
\usepackage{algorithm} 
\usepackage{amsfonts} 
\usepackage{balance} 
\usepackage{algorithmic}
\usepackage{amsmath}

\newcommand{\kl}{D_{\mathrm KL}}
\newcommand{\els}{\mathcal ELS}
\newcommand{\argmin}{\mathop{\arg\min}}
\newcommand{\argmax}{\mathop{\arg\max}}

\newcommand{\abg}{\{\alpha,\beta,\gamma\}}

\begin{document}

\title{qecGPT: decoding Quantum Error-correcting Codes with \\Generative Pre-trained Transformers}

\author{Hanyan Cao}
\affiliation{
CAS Key Laboratory for Theoretical Physics, Institute of Theoretical Physics, Chinese Academy of Sciences, Beijing 100190, China
}
\affiliation{
 School of Physical Sciences, University of Chinese Academy of Sciences, Beijing 100049, China
}
\author{Feng Pan}
\affiliation{
CAS Key Laboratory for Theoretical Physics, Institute of Theoretical Physics, Chinese Academy of Sciences, Beijing 100190, China
}

\author{Yijia Wang}
\affiliation{
CAS Key Laboratory for Theoretical Physics, Institute of Theoretical Physics, Chinese Academy of Sciences, Beijing 100190, China
}
\affiliation{
 School of Physical Sciences, University of Chinese Academy of Sciences, Beijing 100049, China
}

\author{Pan Zhang}
\email{panzhang@itp.ac.cn}
\affiliation{
 CAS Key Laboratory for Theoretical Physics, Institute of Theoretical Physics, Chinese Academy of Sciences, Beijing 100190, China
}
\affiliation{School of Fundamental Physics and Mathematical Sciences, Hangzhou Institute for Advanced Study, UCAS, Hangzhou 310024, China}
\affiliation{International Centre for Theoretical Physics Asia-Pacific, Beijing/Hangzhou, China}

\begin{abstract}
We propose a general framework for decoding quantum error-correcting codes with generative modeling. 
The model utilizes autoregressive neural networks, specifically \textit{Transformers}, to learn the joint probability of logical operators and syndromes. 
This training is in an unsupervised way, without the need for labeled training data, and is thus referred to as \textit{pre-training}.  
After the pre-training, the model can efficiently compute the likelihood of logical operators for any given syndrome, using maximum likelihood decoding.  It can directly generate the most-likely logical operators with computational complexity $\mathcal O(2k)$ in the number of logical qubits $k$, which is significantly better than the conventional maximum likelihood decoding algorithms that require $\mathcal O(4^k)$ computation.

Based on the pre-trained model, we further propose \textit{refinement} to achieve more accurately the likelihood of logical operators for a given syndrome by directly sampling the stabilizer operators. 
We perform numerical experiments on stabilizer codes with small code distances, using both depolarizing error models and error models with correlated noise. 
The results show that our approach provides significantly better decoding accuracy than the minimum weight perfect matching and belief-propagation-based algorithms.
Our framework is general and can be applied to any error model and quantum codes with different topologies such as surface codes and quantum LDPC codes.
Furthermore, it leverages the parallelization capabilities of GPUs, enabling simultaneous decoding of a large number of syndromes.
Our approach sheds light on the efficient and accurate decoding of quantum error-correcting codes using generative artificial intelligence and modern computational power.
\end{abstract}

\maketitle
Quantum computers can potentially solve practical problems which are intractable for classical computers. However, the current implementation of quantum computers has an issue with noise, which limits its power. An essential step towards fault-tolerant quantum computing is quantum error correction (QEC), which now becomes one of the key research frontiers in both theoretical studies and hardware developments~\cite{Panteleev_2022, 10.1145/3519935.3520017} of quantum computation~\cite{google2023suppressing}. In QEC, logical states with $k$ logical qubits are encoded using $n$ physical qubits with redundancy. The effects of continuous errors can be digitalized into a finite set of discrete errors, which can be obtained by measuring the redundant ancilla qubits, giving an error syndrome. Then a decoding algorithm infers the information of errors based on the syndrome and determines an appropriate operation to correct the logical error. However, the decoding problem is a hard problem, for example, it belongs to the class of $\#$P hard problem in the classical error-correcting codes. In quantum codes, decoding is considered to be more challenging than classical code, because the errors inherently degenerate, and the corresponding factor graph for the codes is more complex, for example, in CSS code the factor graph always contains loops with various sizes due to the commutation relations, so standard decoding algorithms such as belief propagation do not work as well as in classical low-density parity check (LDPC) codes.

While a number of algorithms have been proposed for decoding quantum error-correcting codes, we lack general decoding algorithms that are efficient and accurate. The minimum weight perfect matching algorithm~\cite{Dennis2001, higgott2021pymatching} can decode surface code efficiently, however, as a minimum-weight decoder ignores the degeneracy of quantum codes, in principle its performance usually has a gap to the theoretical limit. Moreover, it is less efficient in non-planar graphs and is challenging when applied to code on hypergraphs where the distances between two nodes are not well defined. As a prototype of the maximum likelihood decoder (MLD), tensor network methods (e.g. the boundary matrix product state method~\cite{Bravyi2014}) consider the degeneracies of quantum codes and work close to the theoretical limit in surface code. However, for general codes not defined on lattices with open boundaries, the tensor network contractions are difficult to apply due to the large treewidth of the graph. 
Another issue for existing maximum-likelihood decoders is computing probabilities for $4^k$ logical operators for $k$ logical qubits, which is intractable for a large $k$. 
Moreover, the contraction of tensor networks for each syndrome consumes significantly more computational resources than the minimum-weight decoding algorithms and hence less efficient. Recently, a number of neural network decoders are proposed for leveraging fast inference in neural networks on modern GPUs~\cite{PhysRevLett.119.030501,Varsamopoulos17,Krastanov2017,Varsamopoulos_2020,Overwater2022,Baireuther2017,Davaasuren2020,Gicev2021}.
These methods are based on supervised learning, meaning that training the neural network model requires a large dataset prepared in advance with labels computed using another {teacher} decoding algorithm. Both the dataset size and the accuracy of the teacher decoding algorithm limit the performance of the supervised neural network decoders.

In this work, we propose a maximum likelihood decoding approach based on unsupervised generative modeling in machine learning. The proposed algorithm enjoys efficient decoding using the fast inference of autoregressive neural networks, especially on GPUs, the training directly uses the error model and thus does not require preparing labeled data for training. The autoregressive neural networks can be applied to quantum codes with arbitrary topology, e.g. for general quantum low-density parity check (QLDPC) codes, it also supports directly generating logical operators for an arbitrary number of logical qubits $k$ by reducing the computational complexity from  $\mathcal O(4^k)$ (in computing probabilities of all logical operators in conventional maximum-likelihood decoding) to $\mathcal O(2k)$. In the following text we will first introduce how to link the decoding of stabilizer codes to the generative modeling, then introduce the qecGPT, the pre-trained version of our approach using a specific autoregressive model, casual transformers, then introduce the refinement of decoding accuracy based on the pre-trained model.

\paragraph{{Maximum likelihood decoding---}}
Consider a $[[n,k,d]]$ quantum correction code where a logical state $\ket{\phi}$ with $k$ logical qubits is encoded using a code word $\ket{\psi}$ with $n$ physical qubits. The minimum distance between the code words is $d$. When an error occurs on the state $\psi$, it is considered as an effect of applying an error operator $E$ belonging to the Pauli group $\mathcal P_n=\pm i\{I,X,Y,Z\}$. In the stabilizer formalism~\cite{Gottesman1997, nielsen_chuang_2010}, encoded states are stabilized by some operators $\{s\}$, i.e. $S\ket{\psi}=\ket{\psi}$. The operators form a stabilizer group $\mathcal S=\langle g_1,g_n,\cdots,g_m\rangle$, which is an Abelian sub-group of $\mathcal P_n$, and is generated by $m=n-k$ independent generators.

When an error $E$ occurs, the encoding state $\psi$ may not be stabilized by the stabilizers anymore, this can be tested by measuring the ancilla qubits corresponding to the stabilizer generators, yielding syndrome $\gamma(E)=\{\gamma_1(E),\gamma_1(E),\cdots, \gamma_m(E)\}$, with $\gamma_i(E)=0$ if $g_i$ and $E$ commute and $\gamma_i(E)=1$ if they anti-commute. 
In other words, if the syndrome is not trivial, then the error $E$ must anti-commute with some of the stabilizer generators. If the syndrome is trivial, the error $E$ commutates with all $m$ stabilizer generators, then $E$ is either an element of the stabilizer group or belongs to the logical operators, which is generated by $\mathcal L=\langle l_1^{x},l_1^{z},l_2^{x},l_2^{z},\cdots,l_k^{x},l_k^{z}\rangle$. Here $l_i^x$ and $l_i^z$ denote the logical X and logical Z operators of the $i$'th logical qubits respectively. In addition to $2^m$ stabilizer operators and $4^k$ logical operators, there are still $2^m$ operators that do not commutate with stabilizer generators, they belong to the pure error subgroup $\mathcal E$ which is Abelian and satisfies the commutation relation $e_ig_i=(-1)^\delta_{ij}g_je_i$. The three subgroups introduced above indicate a structure of the Pauli group, which is revealed by the decomposition of the Pauli group $\mathcal P_n=\mathcal E\otimes \mathcal L\otimes \mathcal S$.

Based on the $\{\mathcal E,\mathcal L,\mathcal S\}$ decomposition, we can map an error $E$ to a configuration $\alpha,\beta,\gamma$, and vice versa. Here $\alpha\in\{0,1\}^m$ is the configuration of $m$ stabilizer generators, and each value $\alpha_i$ is determined by the commutation relation between $E$ and the pure error generator $E_i$; $\beta\in\{0,1\}^{2k}$ denotes the configuration for logical $X$ and logical $Z$ operators; $\gamma\in\{0,1\}^m$ is the configuration of $m$ pure error generators, and each value $\gamma_i$ is determined by the commutation relation between $E$ and the stabilizer generators. Using this mapping, we can see the degeneracy of errors, that is, given the logical configuration $\beta$, there are $2^m$ assignments of $\{\alpha\}$ give the same syndrome. So computing likelihood of an logical operator configuration $\beta$ needs to consider all $\alpha$ configurations with 
\begin{align}
p(\beta,\gamma) = \sum_{\alpha} p(\alpha,\beta,\gamma).
\end{align}
In this sense,  we consider the total probability of a coset of the stabilizer sub-group $\mathcal S$, rather than the probability of a single error.

However, there are several challenges for maximum likelihood decoding. The first one is that computing the coset probability is a $\#$P problem, no general exact algorithm exists and approximate algorithms e.g. tensor network contractions are usually time-consuming; the second challenge is that one needs to repeat the computation (by summing over all $\alpha$ configurations) for each syndrome; the third challenge is the exponential computational complexity in the number of logical qubits $k$ because conventionally, one needs to enumerate $4^k$ logical operators, compute their closet probabilities, and find the one with the largest probability.

\paragraph{{Generative maximum likelihood decoding}---}
We propose to solve the challenges of maximum likelihood decoding using a framework based on generative modeling.
First we approximate the joint distribution $p(\alpha,\beta,\gamma)$ using a parameterized variational distribution $q_\theta(\alpha,\beta,\gamma)$ satisfying  
\begin{equation}
    q_\theta(\alpha,\beta,\gamma)=q(\alpha|\mathbf \beta,\gamma)q(\beta|\gamma)q(\gamma).
\end{equation}
Here $q(\alpha|\beta,\gamma)$ is the conditional probability distribution of stabilizer configuration $\alpha$ given the logical operator $\beta$ and the syndrome $\gamma$, $q(\beta|\gamma)$ is the conditional probability of $\beta$ given syndrome. $\theta$ denotes the parameters of the variational distribution $q_\theta(\alpha,\beta,\gamma)$. By learning $\theta$, we make the variational distribution close to the true joint distribution $p(\alpha,\beta,\gamma)$ given by the error model, and make the variational conditional distribution $q(\beta|\gamma)$ close to the true conditional distribution $p(\beta|\gamma)$ which is intractable in general. With an accurate estimate of conditional probabilities, we can evaluate the likelihood of logical operators for all syndromes, and \textit{generate} a configuration of logical operators by sampling $q(\beta|\gamma)$. In other words, the learned variational joint distribution satisfies the condition that the configuration for stabilizer generators $\alpha$ can be traced out automatically. 

\begin{figure}[!htbp]
\centering
\includegraphics[scale=1.1]{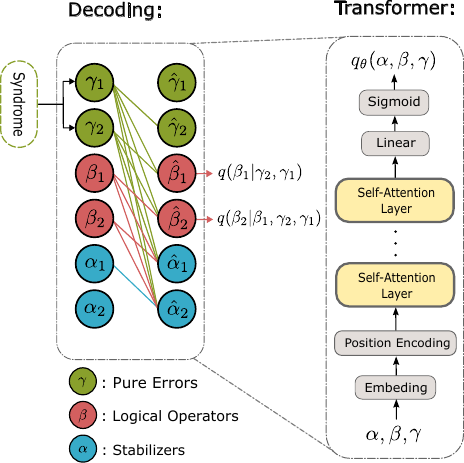}
\caption{Illustration of the structure of qecGPT.} 
\label{fig:qecGPT}
\end{figure}

We further ask the conditional probabilities also satisfy the autoregressive properties for each variable, i.e. $q(\beta|\gamma)=q(\beta_1|\beta_2,...,\beta_{2k}|\gamma)q(\beta_2|\beta_3,...,\beta_{2k}|\gamma)\cdots q(\beta_{2k}|\gamma)$. In this way, all $2k$ logical variables can be generated one by one following the conditional probabilities. This is known as \textit{ancestral sampling} which is an unbiased sampling from the variational conditional distribution $q(\beta|\gamma)$
~\cite{Bishop}. 
The pictorial representation of the generative modeling is illustrated in Fig.~\ref{fig:qecGPT}, where we can see that all the variables are assigned order and each variable only relies on the variables prior to it. I.e., the conditional probability of configuration of a variable $s_i$ is a function of configurations of variables before it $\{s_1,s_2,\cdots,s_{i-1}\}=\mathbf{s}_{<i}$, with $q(s_i|\mathbf{s}_{<i})$.
This property of the parameterization is known as the autoregressive property, also known as \textit{causal} property, if we regard the variables before it as its ``history'', and the variables behind it as its ``future''. Many neural network models satisfy this property and are known as autoregressive neural networks, especially in the models for natural languages where the words are generated one by one.

\paragraph{{The Generative Pre-trained Transformers---}}
In this work, to parameterize $q_\theta(\alpha,\beta,\gamma)$ we adopt the \textsc{Transformers}, one of the most powerful autoregressive neural networks~\cite{Vaswani2017} and has been used in many applications including chatGPT~\cite{ Radford2018ImprovingLU, chatGPT, openai2023gpt4}. We use the decoder layer of the Transformer composed of an embedding layer and a positioning encoding layer which map the input configuration to a higher dimensional feature space, attention layers with triangular mask (to ensure the autoregressive properties), and a linear layer and a sigmoid output layer that outputs the joint probability distributions.
The details of the Transformer can be found in the appendix.

\begin{figure*}[!htb]
\centering
{\includegraphics[scale=0.5]{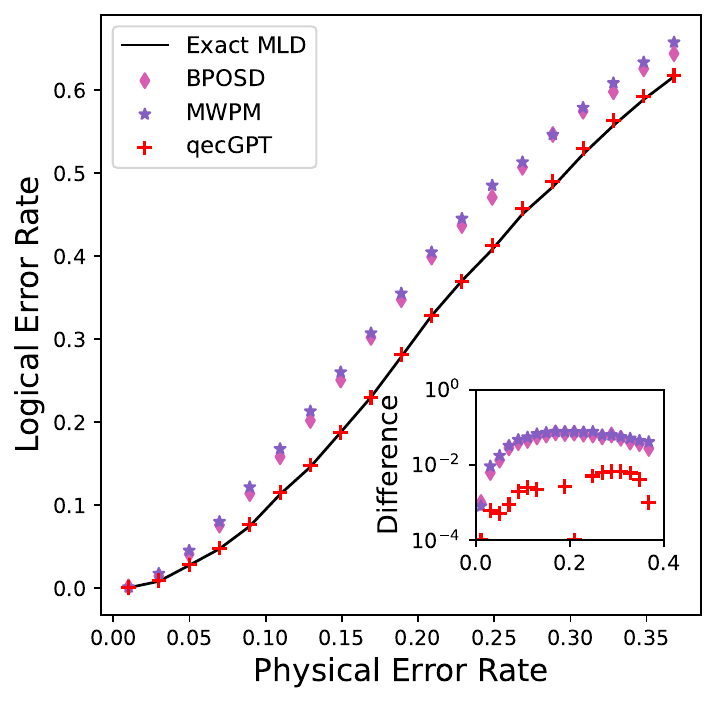}}\hspace{2pt}
{\includegraphics[scale=0.5]{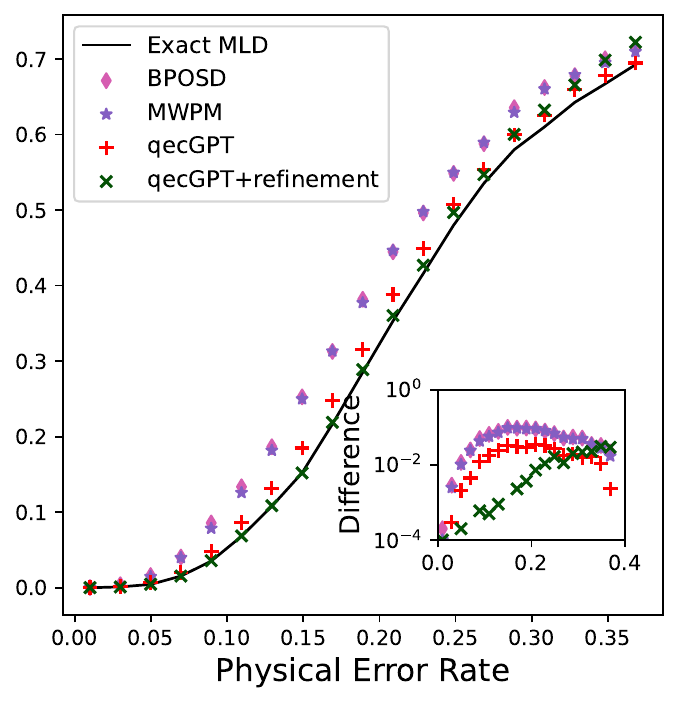}}\hspace{2pt}
{\includegraphics[scale=0.5]{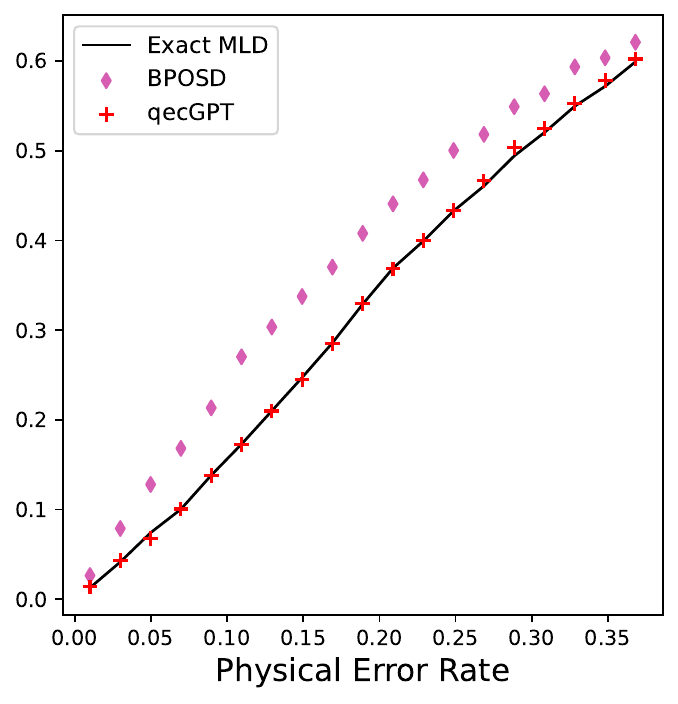}}
\caption{Logical error rates of our algorithm (qecGPT and qecGPT+refinement) with different physical error rates, compared with MWPM, and
BPOSD algorithm on the [13, 1, 3] surface code (left), [41, 1, 5] surface code (middle), and [12, 1, 2] 3D-surface code (right). The error model is the depolarizing model. Each data point in the figures is averaged over $10000$ error instances. The black lines are the optimal maximum-likelihood decoding algorithm which exactly sums all stabilizer configurations. The insets are the difference between the approximate algorithms and the exact algorithm.} 
\label{fig:suface_code}
\end{figure*}

The parameters are learned to minimize the distance between the true distribution 
$p(\alpha,\beta,\gamma)$ and the variational distribution parameterized using the Transformer.
In this work, we assume that we have samples of the noise model $p(E)$, we choose the forward Kullback-Leibler divergence as the distance measure of two probability distributions \begin{align}\kl(p|q)=\sum_{\alpha,\beta,\gamma}p(\alpha,\beta,\gamma)\log\frac{p(\alpha,\beta,\gamma)}{q_\theta(\alpha,\beta,\gamma)}\nonumber.
\end{align}
This yields a negative log-likelihood loss function
\begin{equation}
    \hat\theta = \argmin_\theta \kl = \argmin_\theta \left[-\sum_{\alpha,\beta,\gamma\sim p} \log q_\theta(\alpha,\beta,\gamma)\right]\nonumber.
\end{equation}
And the parameters are updated using a gradient-based optimizer.
After training, given a syndrome $\gamma$, we can generate a configuration of the logical operator $\beta=\{\beta_1,\beta_2,\cdots,\beta_{2k}\}$ one by one using the learned conditional probabilities 
$$\hat \beta_i  = \argmax_{\beta_i} q(\beta_i|\beta_1,\beta_2,\cdots,\beta_{i-1},\gamma_1,\gamma_2,\cdots,\gamma_m).$$
So a logical configuration is generated variable-by-variable given a syndrome. This is analogous to the generation of text from the chatGPT~\cite{Radford2018ImprovingLU,chatGPT,openai2023gpt4}, where the text is generated word-by-word given a prompt. Notice that by training once, a single variational distribution gives the conditional probabilities of all $2^m$ syndromes, maximizing the likelihood of all syndromes. So we term it \textit{pre-training}. The advantage of the pre-training is that the conditional probability for any syndrome can be computed efficiently using a single pass of the neural network. We can further optimize the accuracy of the conditional probability based on the pre-trained model given a particular syndrome, which we term as \textit{refinement}. 
There could be approaches for the refinement, for example, we can minimize the $\kl(q_\theta(\alpha,\beta|\gamma)|p(\alpha,\beta|\gamma))$ using e.g. the method of variational autoregressive networks. 
In this work, we propose a straightforward way for the refinement with a small number of logical qubits $k$, taking advantage of the generative modeling. In addition to generating $\beta$ configurations, we can also generate efficiently the stabilizer configurations $\alpha$ using $q(\alpha,\beta,\gamma)$, and use them to evaluate an unbiased estimate of the joint probability of $\beta$ and $\gamma$
\begin{align} p(\beta,\gamma)&=\sum_{\alpha}q_\theta(\alpha,\beta,\gamma)\frac{p(\alpha,\beta,\gamma)}{q_\theta(\alpha,\beta,\gamma)}\approx\frac{1}{N}\sum_{\alpha\sim q}\frac{p(\alpha,\beta,\gamma)}{q_\theta(\alpha,\beta,\gamma)}\nonumber.
\end{align}
Here we use the samples of the variational distribution and the reweighting to compute an unbiased estimate of the joint distribution, $N$ is the number of samples.

An advantage of our approach is the insensitivity to the topology of the code, i.e. the connectivity of the stabilizer generators, or in other words, the structure of the parity check matrix. The transformer representation of the variational distribution can be used for any code topology without modifying the structure of the transformer, thanks to the self-attention mechanism which can automatically capture correlations in variables.
Our approach is also insensitive to the true parameters used in the error model. For example, under the depolarizing noise model, we can train qecGPT with a particular physical error rate smaller than the threshold, and use our model to decode error under the depolarizing noise model with distinct error rates, without significantly increasing the logical error rate. This effect is quite common in inference problems with mismatched parameters, e.g. in~\cite{Zhang2014}. We refer to the Appendices for details.

\paragraph{{Numerical experiments}---}
We evaluate our algorithm by comparing the logical error rate of our algorithm to the minimum weight perfect matching (MWPM)~\cite{higgott2021pymatching} and belief propagation augmented by ordered statistics decoding (BPOSD)~\cite{Roffe2020} algorithm on the surface code. In Fig.~\ref{fig:suface_code} the surface codes have $k=1$ logical qubits. In Fig.~\ref{fig:suface_code}(left) the distance $d=3$ is small, and we see that the pre-trained model qecGPT performs very close to the exact maximum likelihood decoding which computes the exact likelihood for all $4^k$ logical operators using exact tensor network contractions. In Fig.~\ref{fig:suface_code} middle, the system is larger with $d=5$, and we see that the pre-trained model gives slightly worse results than the exact algorithm while still much better than MWPM and BPOSD. We also see that the refinement significantly improves the performance of qecGPT, making the performance very close to the optimal MLD decoder.

To demonstrate the generality of our approach with different code topologies, we also test the algorithm in the stabilizer code on a 3-dimensional lattice, which is usually termed as \textit{3D surface code}~\cite{PhysRevA.100.012312}. Note that for the 3D structure code, the MWPM algorithm does not apply directly, so we only compare the logical error rates of qecGPT with BPOSD. We can see from Fig.~\ref{fig:suface_code} that the logical error of qecGPT coincides very well with the exact MLD algorithm and significantly outperforms BPOSD. Additional numerical results and performance comparisons with $k>1$ logical qubits and under a noise model with correlated noise can be found in Appendices.

\paragraph{{Discussions}---}
We have introduced a general framework for decoding quantum error correction code with generative modeling. Our method approximates the joint distribution of errors using variational autoregressive neural networks. 
We propose a pre-trained model for the fast generation of the maximum-likelihood logical operators and a refinement to increase the accuracy given a syndrome. 
The advantage of our generative molding is that it solves the difficulties of the maximum likelihood decoding in summing over $2^n$ stabilizer configurations and in computing the probability of $4^k$ logical configurations. Another advantage is its generality in code topologies, e.g. it can be applied to 2D codes and QLDPCs code without modifying the model or the algorithm.

In our work, we have successfully trained the qecGPT using a single GPU and conducted experiments on small codes with distances up to $d=7$. Although the decoding process is fast and efficient, the training phase is slow and poses a challenge when it comes to applying it to a larger code. However, we believe that this bottleneck can be resolved by using a larger model and exploiting more computational resources, such as multiple GPUs or even a supercomputer. This approach is similar to the behavior of chatGPT, which has shown remarkable performance when trained with a large amount of data and computational power, as reported in~\cite{openai2023gpt4,Radford2018ImprovingLU}. We intend to explore this avenue in the future and see how it can further improve the qecGPT's efficiency and scalability.

\begin{acknowledgments}
A \textsc{python} implementation and a Jupyter Notebook tutorial of our algorithm are available at \cite{github}.
    We thank Weilei Zeng, Lingling Lao, and Ying Li for their helpful discussions and Michael Vasmer for providing 3D surface code data. 
\end{acknowledgments}
%
\appendix
\onecolumngrid
\section{Stabilizer codes}
The stabilizer code~\cite{Gottesman1997} is a very important class of quantum error-correcting codes.
Here we will first describe the stabilizer codes and then introduce the decoding algorithms.
Consider a $[n, k, d]$ stabilizer code, states of $k$ logical qubits are encoded to $n$ physical qubits states. The states $\ket{\phi}$ of $n$ physical form a $2^n$ Hilbert space $\mathcal{H}_n$, thus the encoding states $\ket{\psi}$ of $k$ logical qubits form a subspace of $\mathcal{H}_n$ and can be represented by superposition of $\ket{\phi}$. The bit-flip $X$ and phase-flip $Z$ errors may occur on a single qubit state. Then, for $n$ qubits state $\ket{\psi}\in\mathcal{H}_n$, all errors form a group $\mathcal{P}_n = \mathcal{P}^{\otimes n}$ Pauli group. An elements $E\in \mathcal{P}_n$ acting on $\ket{\psi}$ may cause an error state $\ket{\psi'}$. The quantum error correction is to find a recover operator $E'$ to correct the state $E'\ket{\psi'} = \ket{\psi}$. A straightforward idea is to find which error $E$ has occurred and $E' = E$ because of the self-inverse property of Pauli operators. However, a special encoding allows us to find only a collection of operators and any operator belonging to this collection can recover the error. Such an encoding method is called stabilizer code due to the construction is based on a subgroup $\mathcal{S}$ of $\mathcal{P}_n$ called stabilizer group. This group satisfies the following properties:

(a). $\mathcal{S}$ is an abelian group.

(b). $ -I \notin \mathcal{S}$.

Then the encoding states can be chosen as follow:
\begin{equation}
\centering
    \{\ket{\psi} \ \mid\ S\ket{\psi} = \ket{\psi},\ \forall S \in \mathcal{S},\ \forall \ket{\psi}\in \mathcal{H}_n\}
    \label{eq:1}
\end{equation}

According to the properties of $\mathcal{S}$, there are $2^m$ elements in $\mathcal{S}$, where $m$ denotes number of generators $\langle g_1, \cdots ,g_m\rangle$. If there exists a error $E\in \mathcal{P}_n$, one may observe $\gamma(E)$ with length m called error syndrome  
\begin{equation}
\centering
    \gamma(E)_i = \left\{ 
            \begin{split}
            &0, \quad [g_i, E] = 0\\
            &1, \quad \{g_i, E\} = 0\\
            \end{split} 
            \right.
    \label{eq:2}
\end{equation}
The stabilizers are usually described by a parity check matrix $H$ with size $m\times 2n$, where $m=n-k$ denotes the number of stabilizer generators.

\paragraph{{$\mathbf{F_2}$ representaiton}---}

The $\mathbf{F}_2$ representation is an isomorphic map from the Pauli group to itself. Under the $\mathbf{F}_2$ representation, single qubit Pauli operators are represented by two binary numbers:

\begin{equation}
I \rightarrow 00 \quad X\rightarrow 01 \quad Z \rightarrow 10 \quad Y \rightarrow 11
\label{eq:6}
\end{equation}
At this time, any n-qubits Pauli operator is represented by a binary vector with length $2n$~\cite{Gottesman1997}. The group multiplication is just the addition(mod 2) between vectors. And the commutation relation between two operators $A$ and $B$ can be represented by:
\begin{equation}
\centering
    \mathbf{A}\cdot\Lambda\cdot\mathbf{B}^T = \left\{ 
            \begin{split}
            &0, \quad [A, B] = 0\\
            &1, \quad \{A, B\} = 0\\
            \end{split} 
            \right.
    \label{eq:7}
\end{equation}
In Eq.~\ref{eq:7} the bold letters indicate $\mathbf{F}_2$ representations of operators. The dot symbol $\cdot$ denotes matrix multiplication. And the $\Lambda$ is a $2n\times 2n$ matrix $\begin{pmatrix}
    0 & I\\
    I & 0\\
\end{pmatrix}$. Further, the generators of the stabilizer group form a $m\times 2n$ matrix $H$ named parity check matrix.

We have another two subgroups of $\mathcal{P}_n$. One is the pure errors group $\mathcal{E}$. 
The group $\mathcal{E}$ is an Abelian group with $2^m$ elements and all generators $ \langle e_1, \cdots ,e_m\rangle$ satisfy
\begin{equation}
\left\{
\begin{split}
    e_i g_j &= (-1)^{\delta_{ij}} g_j e_i\\
    e_i e_j &= e_j e_i
\end{split}
\right.
\label{eq:3}
\end{equation}

All generators of the pure error group can be stored as another matrix $M_E$ with size $m\times 2n$, satisfying 
\begin{equation}
\begin{split}
    &H\cdot M_E^T = I_{m\times m}\\
    &H\cdot M_E^L = 0.
\end{split}
\label{eq:A2}
\end{equation}
And it can be determined with a matrix $D'$ which is further computed using Gaussian elimination on matrix $D = (H|I_{m\times m})$. $D'$ can be organized as $D'= (A|B)$, where matrix $A$ is a row echelon matrix. And each row of $M_E$ can be solved by these new equations:
\begin{equation}
    A \cdot (M_E)_i^T = B_i
\end{equation}
Since the number of rows of matrix $H$ is less than the number of columns, there are some free variables. For simplicity, we fix these variables to $0$. Moreover, we want these pure error generators to commute to each other as defined in Eq.~\eqref{eq:3}. 
\begin{equation}
    M_E\cdot \Lambda \cdot M_E^T = 0
\end{equation}
This requires that some stabilizer generators acting on these operators are determined. It is equivalent to adding some rows of $H$, corresponding to the stabilizer generators, on rows of $M_E$. The pseudo-code for the whole process is given as follows:
\begin{algorithm}[H]
\caption{Find Pure Errors $M_E$}
\begin{algorithmic}
\label{alg:1}
\REQUIRE Parity Check Matrix $H$ 
\ENSURE $M_E$
\STATE $m\times 2n = |H|$
\STATE $D = (H|I_{m\times m})$
\STATE $D'(A|B) = GE(D)$, where A is a row echelon matrix.
\STATE $M_E \leftarrow$ Each row of $M_E$ can be solved from the equation $\sum_j A_{kj} (M_E^T)_{ji} = B_{ki}$. All free variables are set to 0.
\FOR{$i \in [1, m]$}
    \FOR {$j>i$}
        \STATE$s_{ij} = (M_E)_i\cdot \Lambda \cdot (M_E)_j^T$
        
        \IF{$s_{ij} \neq 0$}
            \STATE $(M_E)_i = (M_E)_i + H_j$
        \ELSE
            \STATE $(M_E)_i = (M_E)_i$
        \ENDIF
    \ENDFOR
\ENDFOR
\RETURN $M_E$
\end{algorithmic}
\end{algorithm}

In addition to the stabilizer group and the pure error group, another sub-group is the logical-operator group $\mathcal{L}$. It  represents the logical errors of logical qubits, is a non-Abelian group, and is generated by $\langle l_{x1}, l_{z1} \cdots, l_{xk}, l_{zk}\rangle $ satisfying
\begin{equation}
\left\{
\begin{split}
    & l_{(x/z)i} g_j = g_j l_{(x/z)i}\\
    & l_{(x/z)i} e_j = e_j l_{(x/z)i}\\
    & l_{xi} l_{zj} = (-1)^{\delta_{ij}} l_{zj} l_{xi}
\end{split}
\right.
\label{eq:4}
\end{equation}

This group has $4\times2^{2k}$ elements, the constant $4$ comes from the overall phase $\{\pm{1}, \pm{i}\}$. However, during the actual error correction process the overall phase is always ignored.
The matrix of generators of the logical subgroup, $M_L$, can be determined given $H$ and $M_E$. As defined in Eq.~\ref{eq:4}, i.e., the $M_L$ is the kernel of matrix $M = \left(\begin{array}{cc}
     & H \\
     & M_E
\end{array}\right)$. 
Here $M$ is a matrix with size $2m\times 2n$, Gaussian elimination of $M$ gives $M'$ and there are $2k$ free variables. The $M_L$ is actually a set of bases of the kernel space, each two rows of the $M_L$ must be linearly independent. Therefore we choose these free variables as a one-hot vector $(0,\cdots,1_i,\cdots,0)$ for $i$th row of $M_L$. And for satisfying the condition Eq.~\ref{eq:4}. We can perform the symplectic Gram-Schmidt orthogonalization procedure (SGSOP) on $M_L$, finding $k$ pairs $L_x$ and $L_z$. The pseudo-code is described in Algorithm~\ref{alg:2}. 
\begin{algorithm}[H]
\caption{Find Logical Operators $M_L$}
\begin{algorithmic}
\label{alg:2}
\REQUIRE $H$, $M_E$
\ENSURE $M_L$
\STATE $M = \left(\begin{array}{cc}
     & H \\
     & M_E
\end{array}\right)$
\STATE $|M|=2m\times 2n$
\STATE $M' = GE(M)$, where $M'$ is a row echelon matrix.
\STATE $M_L \leftarrow$ Each row of $M_E$ can be solved from the equation $M'\cdot (M_L^T)_{i} = 0$. The free variables are set to $(0_1,\cdots,1_i,\cdots,0_{2k})$ for $i$th row of $M_L$.
\FOR{$i \in [1, k]$}
    \FOR {$j \in [i+1, 2k]$}
        \IF{$(M_L)_i\cdot \Lambda \cdot (M_L)_j^T \neq 0$}
        
            \STATE $(M_L)_{i+1} \leftrightarrow (M_L)_j$
            
            \STATE break
        \ENDIF
    \ENDFOR
\ENDFOR

\FOR{$i \in [1, k]$}
    \FOR {$j \in [i+2, 2k]$}
        \IF{$(M_L)_i\cdot \Lambda \cdot (M_L)_j^T \neq 0$}
            \STATE $(M_L)_{j} = (M_L)_j + (M_L)_{i+1}$
        \ELSIF{$(M_L)_{i+1}\cdot \Lambda \cdot (M_L)_j^T \neq 0$}
            \STATE $(M_L)_{j} = (M_L)_j + (M_L)_{i}$
        \ENDIF
    \ENDFOR
\ENDFOR
\RETURN $M_L$
\end{algorithmic}
\end{algorithm}
This is a general algorithm to find logical operators for a given $H$. Actually, one can not distinguish the logical X and logical Z through this algorithm. However for the CSS code, the parity check matrix of which can always be written as $H = (H_z|H_x)$, the $H_x$ and $H_z$ are always treated respectively.
\section{The $\{\els\}$ decomposition}

Note that under the $\mathbf{F_2}$ representation, any two operators are commutative to each other. And the information of anticommutation is stored in a special class of inner products Eq.~\ref{eq:7}. This allows us to use a more efficient way to represent Pauli operators. The Pauli group becomes a self-inverse and Abelian group, under the $\mathbf{F}_2$ representation. Thus any error operator can be generated by generators $\langle e_1, \cdots ,e_m, l_{x1}, l_{z1} \cdots ,l_{xk}, l_{zk}, g_1, \cdots ,g_m\rangle$ and their powers $(\gamma, \beta, \alpha)$. 
\begin{equation}
    E = \prod_{i,j,k} e^{\gamma_i}_i\times l^{\beta_j}_j\times g^{\alpha_k}_k,
\label{eq:8}
\end{equation}
with $\alpha=\{\alpha_k\}\in\{0, 1\}^m,\,\,\, \beta=\{\beta_j\}\in\{0,1\}^{2k},\,\,\, \gamma=\{\gamma_i\}\in\{0,1\}^{m},\,\,\,E\in \mathcal{P}_n$.
This means that there is a correspondence between an error operator $E$ and a $\alpha,\beta,\gamma$ configuration.
\begin{equation}
E \Longleftrightarrow (\alpha,\beta,\gamma)
\end{equation}
We term the power-configuration $(\gamma, \beta, \alpha)$ the $\els$ configuration of an operator which forms a binary vector with length $2n$ $\{\alpha,\beta,\gamma\}^{2n}$. Given an $\els$ configuration, one only needs a series of vector additions (under the $\mathbf{F}_2$ representation) to generate the corresponding operator. On the other hand, given an error operator, the corresponding $\els$ configuration can be determined using Eq.~\ref{eq:7} and 
\begin{equation}
\centering
    \alpha(E)_i = \left\{ 
            \begin{split}
            &0, \quad [e_i, E] = 0\\
            &1, \quad \{e_i, E\} = 0\\
            \end{split} 
            \right.
    \label{eq:9}
\end{equation}

\begin{equation}
\centering
    \beta(E)_{(x/z)i} = \left\{ 
            \begin{split}
            &0, \quad [l_{(z/x)i}, E] = 0\\
            &1, \quad \{l_{(z/x)i}, E\} = 0\\
            \end{split} 
            \right.
    \label{eq:10}
\end{equation}

\begin{equation}
\centering
    \gamma(E)_i = \left\{ 
            \begin{split}
            &0, \quad [g_i, E] = 0\\
            &1, \quad \{g_i, E\} = 0\\
            \end{split} 
            \right.
    \label{eq:11}
\end{equation}

\section{The minimum weight decoder}
Decoding is to determine the recovery operator of the quantum error correction code given the syndrome.  There are basically two kinds of decoding algorithms. The first kind of decoding algorithm is known as the minimum weight decoder, it determines an error with the maximum probability that satisfies all the syndrome constraints. In this sense, when a nontrivial syndrome $\gamma$ has been measured, the minimum weight decoding algorithm finds an error operator $\hat E(\gamma)=\argmax_E P(E(\gamma))$, where $P(E(\gamma)))$ is the generation probability in the error model which is consistent with the syndrome.  The most famous algorithm of the first kind is the Minimum Weight Perfect Matching (MWPM) algorithm, which assigns a weight to each edge in the code graph using the probability of the error in the error model~\cite{higgott2021pymatching}, then finds the shortest error chain given the syndrome, it can be done by employing an efficient algorithm e.g. Blossom algorithm~\cite{Kolmogorov2009}. The MWPM algorithm can decode the surface code efficiently. However, it is challenging for the MWPM algorithm to decode when the code graph is a hypergraph, where each edge of the code graph links to more than two nodes and the distance between two nodes is ill-defined~\cite{higgott2021pymatching}. The main limitation of the minimum weight decoding algorithm is that the error with the maximum probability may not be the right recovery operator because of the degeneracy of the quantum code. 

\section{The maximum likelihood decoder}
The second kind of decoding algorithm is known as the maximum likelihood decoder.
Notice that in the $\els$ decomposition, the stabilizers, and the logical operators do not modify the syndrome $\gamma$. This means that any element of the normalizer of the stabilizer group, $\mathcal N(S)=\mathcal L\otimes \mathcal S$, does not change the syndrome. One can imagine that given an error that produces the syndrome, one can apply, on top of the error, any element from $\mathcal N(s)$ without modifying the syndrome. So in principle, instead of considering a single error that produces the syndrome, one should consider all possible errors that produce the syndrome, which form a closet of the $\mathcal N(S)$. The maximum likelihood decoder determines a logical operator $L$ by considering all operators in the same equivalent class $C(L,\gamma)$ rather than considering a single error, equivalently, it is summing the probabilities of the closet $C(\beta(L),\gamma)$, where $\beta(L)$ is the $\beta$ configuration corresponding to the logical operator $L$. This respects the degeneracy of quantum codes and is the best decoding algorithm one could do. Obviously, the coset probability can be computed by summing probabilities of all elements of the stabilizer group with a particular syndrome and logical operator. The summation can be done by considering all possible $\alpha=\{0,1\}$ configurations.

\begin{align}
    \hat L =\mathop{\arg\max}_{L\in\mathcal{L}} P(C(L,\gamma)) = \sum_{S\in \mathcal{S}} P(E(\gamma)\times L\times S)
    = \sum_{\mathbf{\alpha}} P(e(\gamma)\times L\times(g^{\alpha_1}_1 \times \cdots \times g^{\alpha_m}_m)),
\end{align}
where $\times$ denotes the multiplication of group elements, $e(\gamma)$ is the pure error corresponding to syndrome the Using the representation of $\beta$ configuration for a logical operator, we have 
\begin{align}
    \hat\beta&= \argmax_{\beta} \sum_{\alpha} P(\alpha, \beta, \gamma).
\end{align}
The computation of summing all possible $\alpha$ configurations is analogous to the computation for the partition function of an Ising spin glass, but one needs to do the computation for all possible $\beta$ configurations.
This computation belongs to the \#P problem and there is no exact algorithm to solve it in general in polynomial time. The exact computation of the maximum likely logical operator requires an exponential algorithm. For some special cases, e.g. code in the 2D lattice such as the Surface code, the summation can be approximately computed using the tensor network contractions (e.g. with the boundary matrix product states method) ~\cite{Bravyi2014} of a 2D tensor network constructed for a given syndrome, which could be time-consuming. In addition to the computational cost, the tensor network contraction method has several limitations. The first limitation is that it is difficult to generalize to code on other topologies such as on 3-dimensional lattice, or codes with long-range interactions as in qLDPC, due to the fast increase of the computational cost. and the decrease in accuracy with a topology having large treewidth; the second limitation is that one needs to perform tensor network contractions for each $\beta$ configuration. 

\section{The generative decoder with Transformers}
As described in the main text, our generative decoder models the joint distribution $P(\alpha,\beta,\gamma)$ using the autoregressive neural networks, it factorizes the joint distribution as a product of conditional distributions
\begin{equation}
    q_\theta(\alpha, \beta, \gamma) = q(\alpha | \beta,\gamma) q(\beta | \gamma)q(\gamma)
    \label{eq:13}
\end{equation}
Note that changing the order of the $\alpha,\beta,\gamma$ variables does not change the joint probability distribution due to the Abelian nature of the subgroups. In this work, we always put the logical variable $\beta$ in the middle as in Eq.~\ref{eq:13}. The benefit of this is that the $\beta$ can be always determined prior to the stabilizer variables $\alpha$ in the joint distribution. In this way, decoding is performed by sampling the $\beta$ variables from the marginal distribution
$$q(\beta,\gamma)=\sum_\alpha q_\theta(\alpha,\beta,\gamma).$$

Furthermore, we ask the conditional probabilities are organized in such as way $p(\beta_i | \beta_{j<i})$ that allows generating a $\beta$ configuration among all $4^k$ possible ones variable-by-variable since we have already stored all conditional probabilities for each $\beta_i$ variables.
In this way, we can reduce the computational complexity of generating a maximum-likelihood logical operator out of $4^k$ logical operators to $\mathcal O(2k)$.

\begin{figure*}
    \centering
    \includegraphics[scale=1.0]{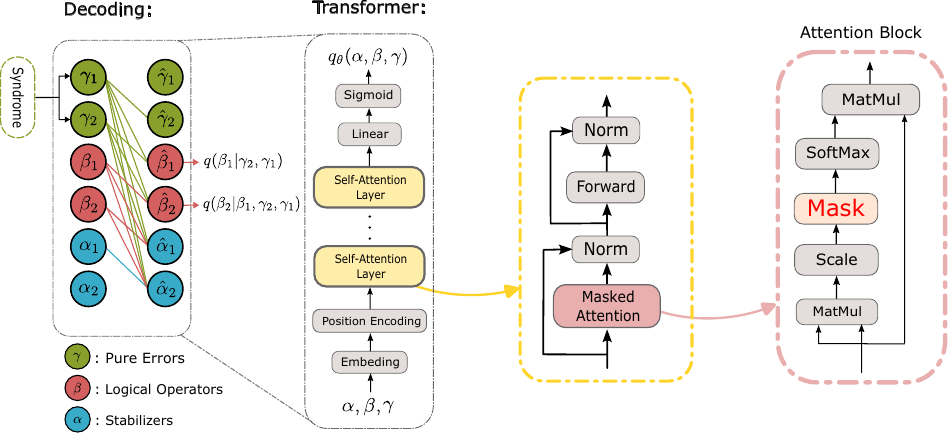}
    \caption{A pictorial illustration of decoding of the quantum error correction code with a generative pretrained transformer.}
    \label{fig:a1}
\end{figure*}

The autoregressive model we used here for representing $q_\theta(\alpha,\beta,\gamma)$ is the encoder part of a Transformer ~\cite{Vaswani2017,fakoor2020trade} with a mask to ensure the autoregressive property. It is also known as causal transformer, the structure is shown in Fig.~\ref{fig:qecGPT}. The input of the Transformer is the configuration $(\alpha, \beta, \gamma)$. The embedding layer increases the dimension of input to the dimension of a model. The information on position will be added and learned from the position encoding. A triangular mask is added in the attention block before the Softmax layer to ensure that each conditional probability of variable $i$ only depends on the variables before $i$ in the input configuration. Multiple transformer encoder layers are added after position encoding, and the final Linear layer maps the data from the model dimension to a length $n$ vector which is the same as the input. The output is a vector $(\hat{\alpha}, \hat{\beta}, \hat{\gamma})$, which uses Sigmoid functions to represent the Bernoulli distributions for the conditional probabilities. For example, as illustrated in Fig.~\ref{fig:qecGPT}, we have 
\begin{align}
\hat\gamma_1 &= \sigma(F_{\gamma_1}(\gamma_1))=q(\gamma_1)\nonumber\\
\hat\gamma_2 &= \sigma(F_{\gamma_2}(\gamma_1))=q(\gamma_2|\gamma_1)\nonumber\\
\hat\beta_1 &= \sigma(F_{\beta_1}(\gamma_1,\gamma_2))=q(\beta_1|\gamma_1,\gamma_2)\nonumber\\
    \hat \beta_2 &= \sigma(F_{\beta_2}(\beta_1,\gamma_1,\gamma_2))=q(\beta_2|\beta_1,\gamma_1,\gamma_2)\nonumber\\
    \hat \alpha_1 &= \sigma(F_{\alpha_1}(\alpha_1,\beta_1,\gamma_1,\gamma_2))=q(\alpha_1|\beta_2,\beta_1,\gamma_1,\gamma_2)\nonumber\\
    \hat \alpha_2 &= \sigma(F_{\alpha_2}(\alpha_2,\alpha_1,\beta_1,\gamma_1,\gamma_2))=q(\alpha_2|\alpha_1,\beta_2,\beta_1,\gamma_1,\gamma_2)
    \label{eq:output}
\end{align}
where $\sigma(\cdot)$ is the Sigmoid function, and function $F$ denotes the map of the neural network. 
We can see that the product of all the output of the transformer gives the joint distribution, as
\begin{align}
    \hat\alpha_2\hat\alpha_1\hat\beta_2\hat\beta_1\hat\gamma_2\hat\gamma_1 &= q(\alpha_2|\alpha_1,\beta_2,\beta_1,\gamma_2,\gamma_1)q(\alpha_1|\beta_2,\beta_1,\gamma_2\gamma_1)q(\beta_2|\beta_1,\gamma_2,\gamma_1)q(\beta_1|\gamma_2,\gamma_1)q(\gamma_2|\gamma_1)q(\gamma_1) \nonumber\\
    &= q(\alpha_2,\alpha_1,\beta_2,\beta_1,\gamma_2,\gamma_1).
    \end{align}
Another important property (but not obvious) that we can obtain from the product of the output of the Transformer is the marginal distribution.
\begin{align}
    \hat\beta_2\hat\beta_1&= q(\beta_2|\beta_1,\gamma_2,\gamma_1)q(\beta_1|\gamma_2,\gamma_1) \nonumber\\
    &= q(\beta_2,\beta_1|\gamma_2,\gamma_1). 
    \end{align}
Which gives the normalized conditional probability of logical variables given a syndrome, and surprisingly ignores the stabilizer configurations $\alpha$. Based on this conditional probability one can evaluate the likelihood of the logical operators and perform the decoding after the neural network is well trained.

\section{Pre-training of the model}
The training of the neural network is conducted by minimizing the distance between the error distribution (given by the error model) and the variational distribution $q_\theta$. Here we adopt the Kullback-Leibler divergence.
\begin{equation}
    \hat \theta = \argmin_\theta D_{\mathbf{KL}}\left[P(\alpha, \beta, \gamma)\ ||\ q_{\theta}(\alpha, \beta, \gamma)\right]
\label{eq:16}
\end{equation}
Here we use the forward KL divergence because we always assume that we have $N$ samples $\{\alpha,\beta,\gamma\}\sim P(\alpha,\beta,\gamma)$ of the errors, which can be obtained by sampling the error model or collected from experiments. Then the loss function can be evaluated as
\begin{align}
\hat \theta &= \argmin_\theta D_{\mathbf{KL}}\left[P(\alpha, \beta, \gamma)\ ||\ q_{\theta}(\alpha, \beta, \gamma)\right]\nonumber\\ 
&=\argmin_\theta \sum_{\alpha,\beta,\gamma}P(\alpha, \beta, \gamma)\left[\log P(\alpha, \beta, \gamma)-\log q_{\theta}(\alpha, \beta, \gamma)\right]\nonumber\\ 
&=\argmin_\theta \frac{1}{N}\sum_{\{\alpha,\beta,\gamma\}\sim P(\alpha,\beta,\gamma)}\left[\log P(\alpha, \beta, \gamma)-\log q_{\theta}(\alpha, \beta, \gamma)\right]\nonumber\\ 
&=\argmin_\theta -\sum_{\{\alpha,\beta,\gamma\}\sim P(\alpha,\beta,\gamma)}\log q_{\theta}(\alpha, \beta, \gamma)\nonumber\\ 
&=\argmin_\theta F_{\mathrm{NLL}},
    \label{eq:17}
\end{align}
where $F_{\mathrm{NLL}}=-\sum_{\{\alpha,\beta,\gamma\}\sim P(\alpha,\beta,\gamma)}\log q_{\theta}(\alpha, \beta, \gamma)$ is the so-called negative log-likelihood loss function or the cross-entropy loss.

In this work, we consider error models that are easy to sample. For example, for the depolarizing model, the errors on each qubit are generated independently. We also consider the correlated noise where the errors are generated according to some pairwise correlations. In this case, we can adopt the Metropolis-Hasting algorithm to sample the error. Once an error operator is sampled, according to the isomorphic mappings between the Pauli group and its ELS representation (i.e. with Eq.~\ref{eq:9}, Eq.~\ref{eq:10}, and Eq.~\ref{eq:11}), a $\abg$ configuration is generated as training data for training the model for all syndromes. So we call it Pre-training, it learns a joint distribution for all syndromes, and also keeps the conditional probabilities $q(\beta|\gamma)$ for all syndromes. 

So the decoding is very fast because one can evaluate the conditional probabilities for each $\beta_i$ variable by the forward passes of the neural network especially using GPUs. Here we describe it in detail using the example of Fig.~\ref{fig:a1}. 
First, a syndrome $\gamma_1,\gamma_2$ is sent as an input to the Transformer, after one forward pass of the Transformer $F_{\beta_1}(\gamma_2,\gamma_1)$, we compute a conditional probability $q(\beta_1|\gamma_2,\gamma_1)$ using e.g. Eq.~\ref{eq:output} and sample a configuration of $\beta_1$ which maximize the conditional probability with 
\begin{equation}
    \hat \beta_1=\argmax_{\beta_1}q(\beta_1|\gamma_2,\gamma_1).\label{eq:beta_argmax}
\end{equation}
Then, we send $\beta_1$ and $\gamma$ as input to the transformer, compute a conditional probability $q(\beta_2|\beta_1,\gamma_2,\gamma_1)$, then sample a $\beta_2$ configuration according to the conditional probability

\begin{equation}
\hat \beta_2=\argmax_{\beta_2}q(\beta_2 | \beta_1,\gamma_2,\gamma_1).
\end{equation}

\begin{figure*}[!htbp]
\centering
    \includegraphics[scale=0.8]{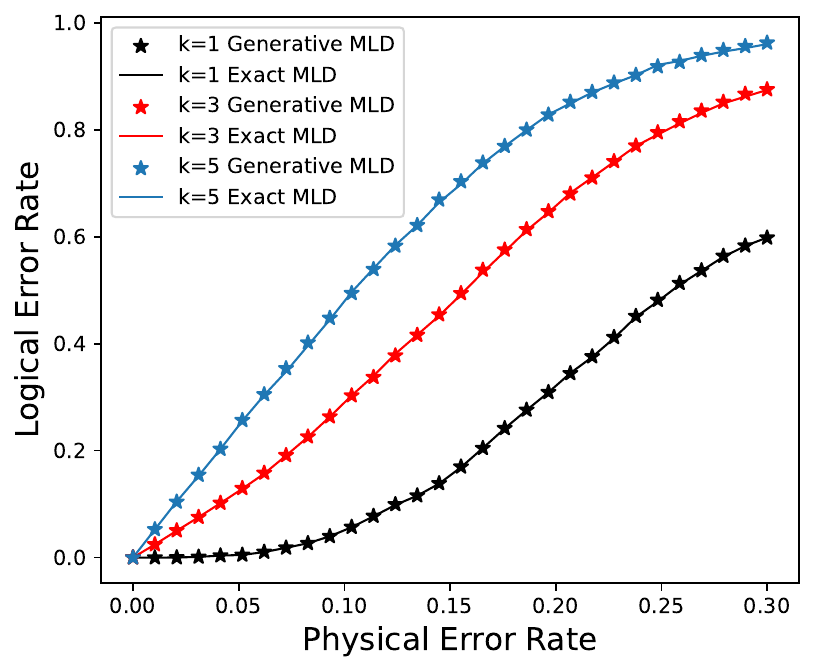}
    \caption{Comparison between the exact maximum likelihood decoding (MLD) and an exact MLD (solid lines) and generative MLD (stars). The coset probabilities $p(\mathbf{\beta}|\gamma)$ and the conditional coset probabilities $p(\beta_i|\beta_{j<i}, \gamma)$ are calculated through contraction of tensor networks. We do the lattice surgeries on d=5 Surface Code. The numbers of logical qubits are [1, 3, 5]. Each data point is averaged over $10000$ random syndrome instances.} 
\label{fig:decimation}
\end{figure*}

We call the method of sample $\beta$ variables one by one the \textit{generative MLD decoding}. In Fig.~\ref{fig:decimation}, we compare the logical error rate given by the exact MLD decoding which enumerates all possible $4^k$ logical operators,  and the generative MLD and we can see that on the surface code and we can see that the generative results are almost identical to the exact MLD results while the computational complexity has decreased from $4^k$ to $2k$ for $2k$ conditional probabilities.

We remark that the decoding for multiple syndromes can be done simultaneously because the Transformer can accept a batch of syndromes as an input and process the batched forward pass efficiently, especially using GPUs.

\section{Refinement} 
The pre-training learns a joint distribution $q_\theta(\alpha,\beta,\gamma)$ for all $2^m$ $\gamma$ configurations and offers a fast forward pass for decoding. For a given syndrome, we can spend more computational cost to further enhance the accuracy of decoding, we term it as refinement. 
A straightforward way for the refinement of the transformer is minimizing the distance between the conditional distribution and the true conditional distribution. However, since it is not possible to obtain samples of the conditional distribution $p(\alpha, \beta| \gamma)$ given a syndrome directly from the error model due to the lack of the normalization factor, we can not directly refine the variational distribution given the syndrome using the forward KL divergence $\kl(p(\alpha,\beta|\gamma) \|q(\alpha,\beta|\gamma)$. Instead, we can do the refinement by minimizing the reverse KL
\[\kl(q(\alpha,\beta|\gamma) \|p(\alpha,\beta|\gamma)),\]
because the variational distribution $q(\alpha,\beta|\gamma)$ is always sample able. However, using the reverse KL divergence requires computing the gradients using the samples of the variational distribution with the REINFORCE method~\cite{williams1992simple} (also known as policy gradients), analogous to reinforcement learning. The procedure is quite similar to the 
variational autoregressive neural networks for minimizing the variational free energy for statistical mechanics problems~\cite{PhysRevLett.122.080602}.

However, minimizing the backward KL is computationally expansive. In this work, we propose another way for the refinement of qecGPT, which we call the \textit{generative refinement}. The idea is to explicitly compute the summation of the stabilizer configurations using the $\alpha$ configurations sampled from the variational distribution (taking advantage of the generative models), given a syndrome configuration $\gamma$. That is the unbiased version of the joint probability $p(\beta,\gamma)$ can be computed using samples of stabilizer configurations that are reweighted.  

\begin{align} p(\beta,\gamma)&=\sum_{\alpha}q_\theta(\alpha,\beta,\gamma)\frac{p(\alpha,\beta,\gamma)}{q_\theta(\alpha,\beta,\gamma)}\approx\frac{1}{N}\sum_{\alpha\sim q}\frac{p(\alpha,\beta,\gamma)}{q_\theta(\alpha,\beta,\gamma)}\nonumber.
\end{align}
On the R.H.S. of the last equation, we use the samples of the variational distribution and the reweighting to compute an unbiased estimate of the joint distribution. 
The samples of $\alpha$ configurations can be computed using conditional probabilities, but different from the way we obtained $\beta$ configurations. Again we use Fig.~\ref{fig:a1} as a simple example. Suppose we are given a syndrome configuration $\gamma_1,\gamma_2$ and we want to estimate the joint probability of a $\beta$ configuration $\beta_1,\beta_2$ and the syndrome configuration $p(\beta_2,\beta_1,\gamma_2,\gamma_1)$. We can send $\beta_2,\beta_1,\gamma_2,\gamma_1$ as an input to the transformer, compute the conditional probability 
and sample the $\alpha_1$ configuration according to this probability distribution \[\alpha_1\sim q(\alpha_1|\beta_2,\beta_1,\gamma_2,\gamma_1).\]  Notice that this is different from the sampling procedure to determine the $\beta$ configuration in the decoding of the pre-trained model, as here we are sampling from the distribution while in decoding of the pre-trained model maximizing the conditional probabilities as shown in Eq.~\eqref{eq:beta_argmax}. After determining the value of $\alpha_1$, we send $\alpha_1,\beta_2,\beta_1,\gamma_2,\gamma_1$ as an input to the transformer, compute the conditional probability and sample the $\alpha_1$ configuration according to this it \[\alpha_2\sim q(\alpha_2|\alpha_1,\beta_2,\beta_1,\gamma_2,\gamma_1).\] 
The advantage of the reweighting formula is the unbiased estimates, and it greatly improves the decoding accuracy from the pre-training, but the drawback is that we have to evaluate $4^k$ configurations of $\beta$ variables for the maximum-likelihood decoding. 

\section{Fast decoding with mismatched training parameters to the error model} 
In maximum likelihood decoding, computation of the likelihood of logical operators requires the parameters of the error model. In the qecGPT, the parameters of the error model explicitly appear in the (slow) training process but do not appear in the (fast) decoding process. It would be very efficient if a transformer trained using a set of parameters of the noise model but used for decoding under another noise model parameter. 
In the area of statistical inference, this is known as inference with mismatched parameters and it is well known that although not optimal, sometimes the mismatched parameters already provide very accurate inference results. For example, in the community detection problem (which can be studied using the statistical inference of the stochastic block model), It has been shown that the inference using the parameters at the phase transition always gives good results, and is even optimal in the sense of the range of detection~\cite{Zhang2014}. Inspired by the results of ~\cite{Zhang2014}, we can always train the qecGPT using the parameters of the noise model at the theoretical phase transition point. To give a concrete example, for the surface code, the phase transition happens with $n\to\infty$ at a physical error rate $p\approx 0.189$~\cite{Bombin_2012}. We can train the qecGPT with $p=0.189$ and decode for codes with other physical error rate values.

\begin{figure}
    \centering \includegraphics[scale=0.8]{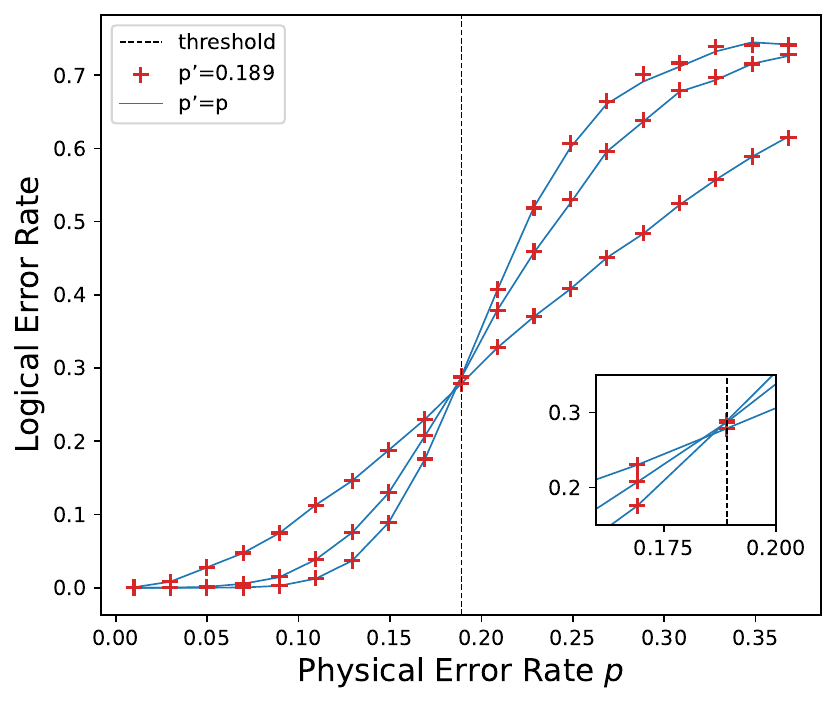}
    \caption{The performance of exact maximum likelihood decoding with mismatched parameters on the $d=[3, 7, 11]$ surface codes under the depolarizing error model. The decoding is implemented using exact tensor network contractions with noise parameter $p'$. The blue line shows the decoding results with $p'=p$, i.e. with matched error model parameters. The red symbols show the decoding results with $p'=0.189$, which is fixed to the physical error rate at the threshold (phase transition). Each data point is averaged over $10000$ random syndrome instances.}
    \label{fig:mismatch}
\end{figure}

In Fig.~\ref{fig:mismatch}, we have tasted the performance of mismatched decoding of the surface code with various code distances. The maximum likelihood decoding is performed using exact tensor network contractions with noise parameter $p'$. We have tested two kinds of $p'$ values. The first one is the matched parameters, where $p'$ is set to the true physical error rate $p$ in the depolarizing noise model; the second one is fixed to $p'=0.189$ which is the threshold (phase transition) of the surface code with $d=\infty$. From the figure, we can see that the results of the MLD decoding with the threshold parameter is indistinguishable from the exact MLD decoding with true parameters.

\subsection{Compared with other neural network decoders}
Recently, several decoding algorithms based on neural networks have been proposed. These include algorithms using Boltzmann machines~\cite{PhysRevLett.119.030501}, multilinear neural networks~\cite{Varsamopoulos17, Krastanov2017, Varsamopoulos_2020, Overwater2022}, the Long Short-Term Memory (LSTM) neural networks~\cite{Baireuther2017}, and the convolutional neural network (CNN)~\cite{Davaasuren2020, Gicev2021}. 
We notice that all of the existing neural network decoders are based on supervised learning. This means that the training of the models requires a training dataset with labels prepared using another decoder. The labels are either the operators that are computed using the minimum weight perfect matching algorithm, or the correct type of logical operators that are computed using a maximum likelihood decoding algorithm. For detailed introductions to the neural network decoders we refer to~\cite{Battistel2023RealTimeDF}.

In this sense, our generative decoders are much different from the existing neural network decoders. Among many differences, the most crucial difference is that our approach uses unsupervised learning rather than supervised learning. It models the joint distribution of errors using neural networks, rather than learning the probability of labels. In other words, for training the neural networks, we do not need to prepare any labeled data. Instead, we directly draw unlabelled samples from the noise model as training data. Moreover, the existing neural decoders belong to \textit{discriminative} learning while our approach belongs to \textit{generative} learning which generates the maximum likely logical operators variable by variable, in analogous to generating a sentence word by word.
Another significant difference between our approach and the existing neural network decoder is that our generative decoder can generate maximum likely logical operators variable by variable, hence is capable of decoding with a large number of logical qubits. On the opposite, the maximum likelihood decoders based on supervised learning require labeled data with $4^d$ different kinds of labels, which is intractable. For example, with $k=7$, our generative decoder can generate logical operators with computational complexity $O(14)$, while the neural network decoders based on the generative learning require $4^7=16384$ different types of labels and hence is intractable.

\section{Additional numerical results}
\begin{figure}[!htbp]
    \centering
	{\includegraphics[scale=0.73]{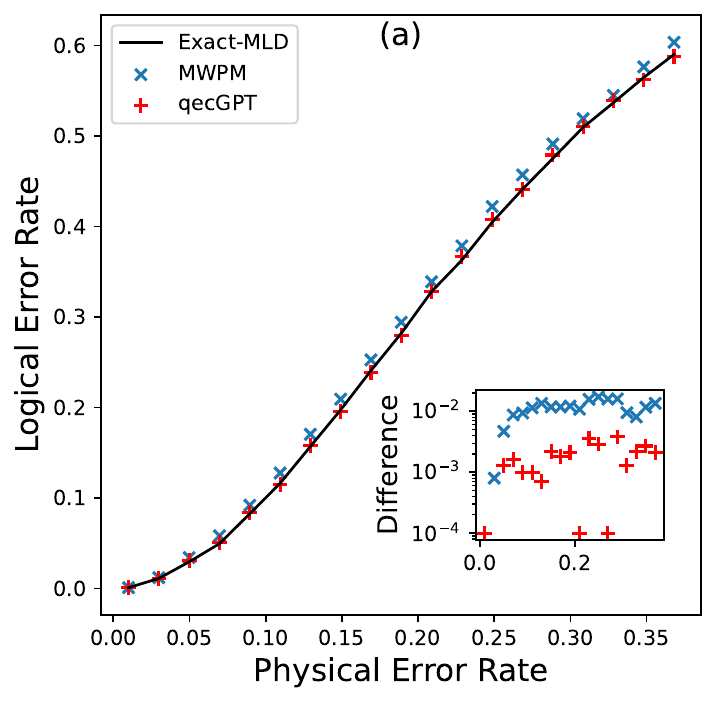}}
	{\includegraphics[scale=0.73]{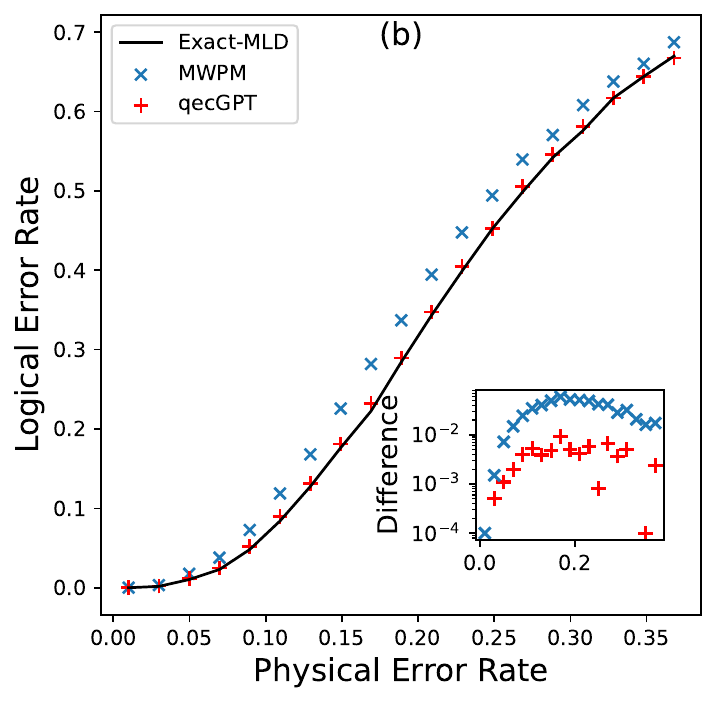}}
	{\includegraphics[scale=0.78]{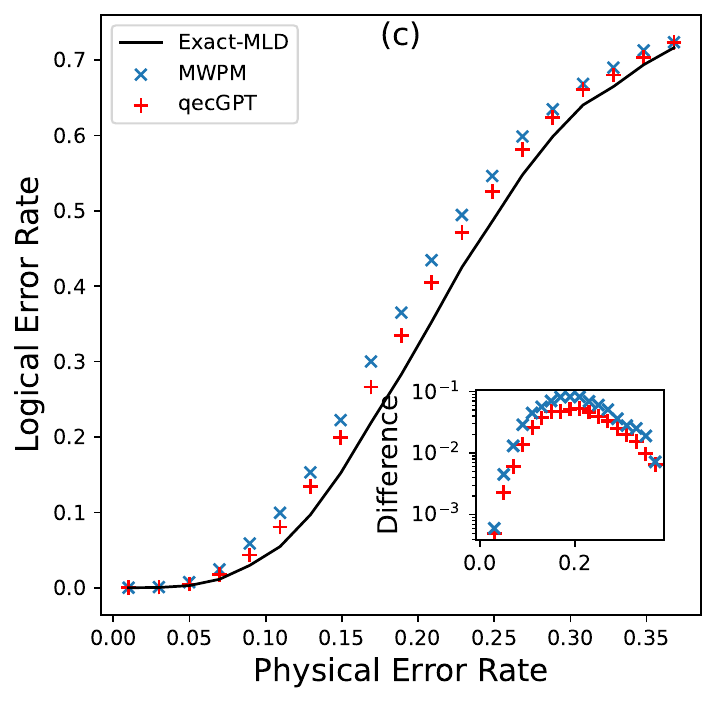}}
\caption{Comparison of decoding performance of various algorithms on the rotated surface code with $d=3$(a), $d=5$(b), and $d=7$(c). Each data point is averaged over $10,000$ random instances. MWPM means minimum weight perfect matching algorithm, and the Exact-MLD is implemented by summing all possible stabilizer configurations using exact tensor network contractions.} 
\label{fig:4}
\end{figure}

In this section, we provide numerical results in addition to the results we have shown in the main text.
\paragraph{\textbf{Rotated Surface Codes}---} 
Here we benchmark qecGPT on the Rotated Surface Codes~\cite{Bombin2007}. This type of code has the same threshold as the surface Code but with only $d^2$ physical qubits for encoding a single logical qubit, rather than $n=d^2+(d-1)^2$ qubits in the surface code. Recently this code is frequently used in quantum hardware experiments~\cite{google2023suppressing}. We compared the performance of qecGPT with MWPM in Fig.~\ref{fig:4} on the rotated surface code with different code distances. It can be seen that qecGPT always provides more accurate results (with lower logical error rates) than MWPM. 

\begin{figure}[!htbp]
    \centering
    {\includegraphics[scale=0.75]{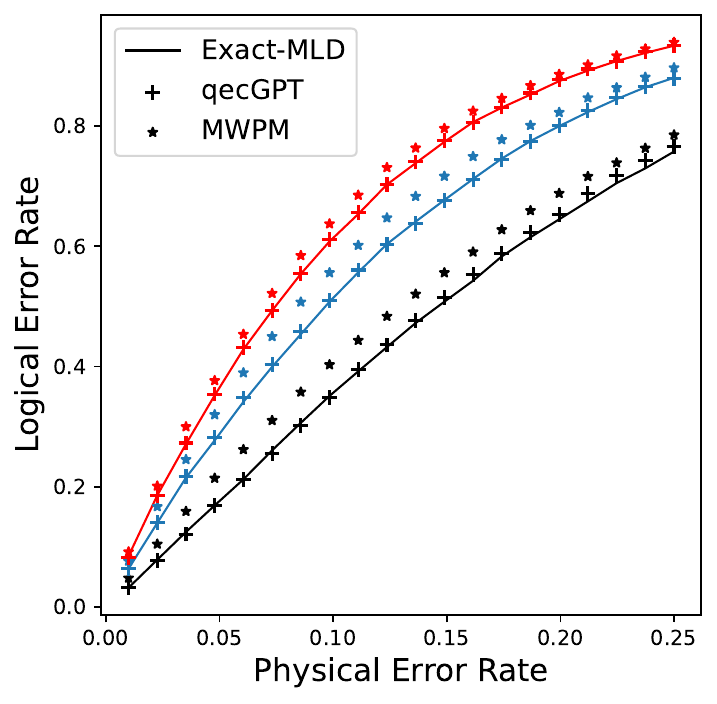}}
   {\includegraphics[scale=0.75]{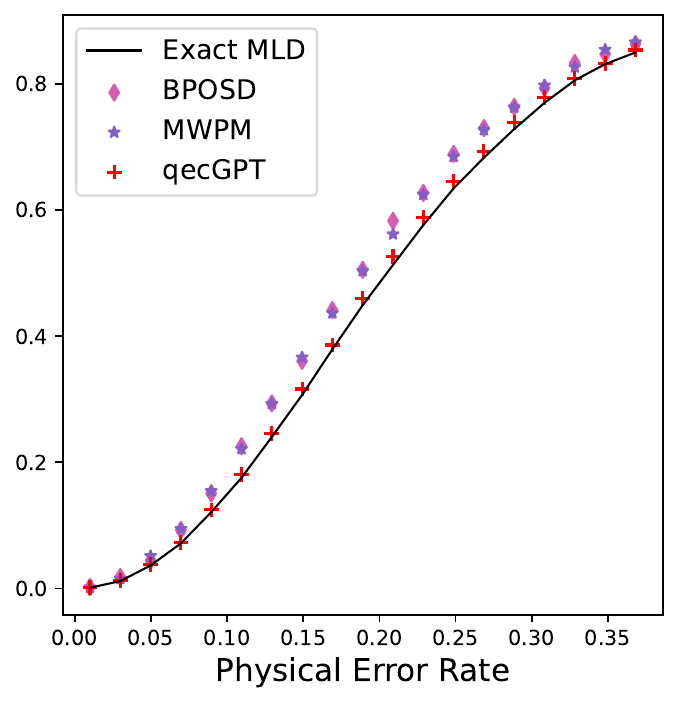} }
\caption{Comparison of decoding performance with $k > 1$  logical qubits.(left) distance-3 Surface Codes with random lattice surgeries. The number of physical qubits is 26 and the numbers of logical qubits are k=3 (black), k=5 (blue), and k=7 (red) respectively.  (right) Toric Code with code distance $d=3$ and $k=2$ logical qubits. In the figures, MWPM is the minimum weight perfect matching~\cite{higgott2021pymatching}, and BPOSD is belief propagation augmented by ordered statistics decoding (BPOSD)~\cite{Roffe2020} algorithm. Exact MLD is the maximum likelihood decoding which sums all possible stabilizer configurations using exact tensor network contractions.
\label{fig:multi}}
\end{figure}

\paragraph{\textbf{With $k >1$ logical qubits}---} An advantage of qecGPT is that it can decode with $k\gg 1$ logical qubits efficiently via generative capability. In maximum likelihood decoding, one usually needs to compare the likelihood of all $4^k$ logical operators with $k$ logical qubits, which is intractable for a large $k$. Instead, the complexity of qecGPT is only $O(2k)$ because it generates logical configurations variable by variable. To evaluate the performance of qecGPT with $k>1$ logical qubits, we did lattice surgery on $d=3$ surface code to increase the number of logical qubits $k$. In detail, we have removed 2 (4, 6) stabilizers randomly and increased the number of logical qubits to 3 (5, 7) respectively. This increases the number of logical operators to $64$ with $k=2$, $1024$ with $k=5$, and $16384$ with $k=7$. During the training of the transformer, we always fix the physical error rate to $p'=0.15$ and use the model to decode with other physical error rates $p$. From Fig.~\ref{fig:multi} we can see that qecGPT outperforms minimum weight perfect matching on both surface code and toric code~\cite{Dennis2001}  with multiple logical qubits. 

\begin{figure}[!htbp]
    \centering
	\includegraphics[scale=0.7]{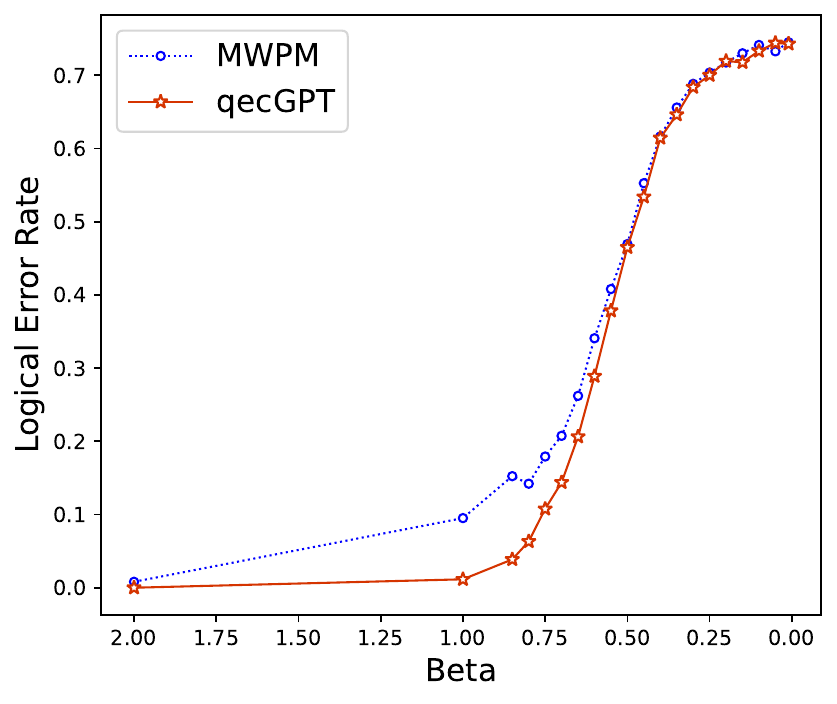}
\caption{Decoding of the surface code with $d=5$ with correlated errors described in the text. The weights of MWPM are determined using the marginal probabilities calculated using samples. Each data point is averaged over $10000$ instances.}
\label{fig:cor}
\end{figure}

\paragraph{\textbf{Correlated noise}---} Although simple error models e.g. depolarizing models assume that errors are independent on each qubit, in practical quantum hardware there inevitably exist correlations between errors on different qubits. In our approach, we only need samples from the error models for training qecGPT, so we can decode error-correcting codes with correlated noise in exactly the same way as we described and evaluated with the independent noise models. In this section, we propose a simple noise model to evaluate our approach for the correlated noise. For a $[n, k, d]$ quantum code, we generate an Ising model on $degree=4$ regular random graph with $n$ spins. The couplings $J_{ij}$ of the Ising model are sampled from a uniform distribution $\mathcal{U}(0, 1)$. And for breaking the $Z_2$ symmetry, a small external field $h=0.3$ is added to Hamiltonian as Eq.~\ref{eq:19}.
\begin{equation}
    H = -\beta\sum_{<ij>} J_{ij} s_i s_j - h\sum_{i} s_i
\label{eq:19}
\end{equation}
Then we draw samples from the Boltzmann distribution using the Metropolis-Hasting algorithm. The error that occurred is determined by a sample $s$. If $s_i=1$, an identity $I$ acts on $i$th physical qubit. Else, There is a Pauli error $\{X, Y, Z\}$ with the same probabilities. We can see that the error configurations are mapped from the configuration of Ising models which are generated from the Boltzmann distribution with long-range correlations. With $\beta=0$, all the configurations are drawn randomly, each spin has a probability of $0.5$ to take $1$ or $-1$, which is analogous to the depolarizing error model with a high physical error rate. With $\beta>0$ the correlations between the error configurations are long-range and non-trivial. With $\beta=\infty$, the ground state of the Ising model is a ferromagnetic configuration, so there are almost no Pauli error appears, corresponding to a low physical error rate.
In Fig.~\ref{fig:cor} we compare the performance of qecGPT with the minimum weight perfect matching algorithm which determines the weights using the marginal probabilities calculated using the samples. In the figure, with $\beta=0$, the error model is purely random so the logical error rate of both MWPM and qecGPT is high. With $\beta$ large, the physical error rate is small and the decoding is very easy we also see that both MWPM and qecGPT give a very low logical error rate. In the middle when the physical error rate is moderate, we can see that qecGPT significantly outperforms the minimum weight perfect matching algorithm.

\subsection{Parameters of the neural networks}
In the numerical experiments, all neural networks were trained on a single NVIDIA A100 GPU. 
To ensure that the distribution of the Transformers has the autoregressive property, an upper-triangle mask is added to the attention block as illustrated in Fig.~\ref{fig:a1}. We note that in the Softmax layer of the Transformers, we have added a mask matrix with diagonal elements set to $1$.
\begin{equation}
    \left(\begin{matrix}
         1 & $-$\infty & \cdots & $-$\infty \\
         \vdots & \ddots & \ddots & \vdots\\
         \vdots &   &\ddots & $-$\infty\\
         1 & \cdots & \cdots & 1\\
    \end{matrix}\right)
\end{equation}
In this setting, the autoregressive property is not satisfied because the $i$th output depends on the $i$th input. The solution is to introduce a virtual variable $x_0$ and modify the order of input variables as
\begin{equation}
    \mathbf{X}(x_1, \cdots, x_n) \rightarrow \mathbf{X}(x_0, \cdots, x_{n-1}),
\end{equation}
and $x_0$ always equals 1.
The hyperparameters for training the Transformers are detailed in Tab.~\ref{tab:1}. Where $D$ is the dimension of the model, $N_h$ is the number of heads of multi-head attention, $N_l$ is the number of encoder layers, and $D_f$ is the dimension of feed-forward layers.
\begin{table}[!htbp]
    \centering
    \begin{ruledtabular}
    \begin{tabular}{cccccccc}
    \textrm{}&
    \textrm{BATCH}&
    \textrm{EPOCH}&
    \textrm{LR}&
    \textrm{$D$}&
    \textrm{$N_h$}&
    \textrm{$N_l$}&
    \textrm{$D_f$}\\
    Sur3 & $10^4$ & $10^5$ & $10^{-3}$ & 256 & 4 & 2 & 256\\
    Sur5 & $10^4$ & $2\times10^5$ & $10^{-3}$ & 256 & 4 & 3 & 256\\
    RSur3 & $10^4$ & $10^5$ & $10^{-3}$ & 256 & 4 & 2 & 256\\
    RSur5 & $10^4$ & $2\times10^5$ & $10^{-3}$ & 256 & 4 & 3 & 256\\
    RSur7 & $10^4$ & $3\times10^5$ & $10^{-3}$ & 512 & 4 & 2 & 512\\
    3DSur2 & $10^4$ & $10^5$ & $10^{-3}$ & 256 & 4 & 2 & 256\\
    Tor3 & $10^4$ & $10^5$ & $10^{-3}$ & 256 & 4 & 2 & 256\\
    N13k3 & $10^4$ & $10^5$ & $10^{-3}$ & 256 & 4 & 2 & 256\\
    N13k5 & $10^4$ & $10^5$ & $10^{-3}$ & 256 & 4 & 2 & 256\\
    N13k7 & $10^4$ & $10^5$ & $10^{-3}$ & 256 & 4 & 2 & 256\\
    \end{tabular}
    \end{ruledtabular}
    \caption{Parameters of qecGPT and hyperparameters in the training.}
    \label{tab:1}
\end{table}


\begin{thebibliography}{32}%
\makeatletter
\providecommand \@ifxundefined [1]{%
 \@ifx{#1\undefined}
}%
\providecommand \@ifnum [1]{%
 \ifnum #1\expandafter \@firstoftwo
 \else \expandafter \@secondoftwo
 \fi
}%
\providecommand \@ifx [1]{%
 \ifx #1\expandafter \@firstoftwo
 \else \expandafter \@secondoftwo
 \fi
}%
\providecommand \natexlab [1]{#1}%
\providecommand \enquote  [1]{``#1''}%
\providecommand \bibnamefont  [1]{#1}%
\providecommand \bibfnamefont [1]{#1}%
\providecommand \citenamefont [1]{#1}%
\providecommand \href@noop [0]{\@secondoftwo}%
\providecommand \href [0]{\begingroup \@sanitize@url \@href}%
\providecommand \@href[1]{\@@startlink{#1}\@@href}%
\providecommand \@@href[1]{\endgroup#1\@@endlink}%
\providecommand \@sanitize@url [0]{\catcode `\\12\catcode `\$12\catcode
  `\&12\catcode `\#12\catcode `\^12\catcode `\_12\catcode `\%12\relax}%
\providecommand \@@startlink[1]{}%
\providecommand \@@endlink[0]{}%
\providecommand \url  [0]{\begingroup\@sanitize@url \@url }%
\providecommand \@url [1]{\endgroup\@href {#1}{\urlprefix }}%
\providecommand \urlprefix  [0]{URL }%
\providecommand \Eprint [0]{\href }%
\providecommand \doibase [0]{https://doi.org/}%
\providecommand \selectlanguage [0]{\@gobble}%
\providecommand \bibinfo  [0]{\@secondoftwo}%
\providecommand \bibfield  [0]{\@secondoftwo}%
\providecommand \translation [1]{[#1]}%
\providecommand \BibitemOpen [0]{}%
\providecommand \bibitemStop [0]{}%
\providecommand \bibitemNoStop [0]{.\EOS\space}%
\providecommand \EOS [0]{\spacefactor3000\relax}%
\providecommand \BibitemShut  [1]{\csname bibitem#1\endcsname}%
\let\auto@bib@innerbib\@empty
\bibitem [{\citenamefont {Panteleev}\ and\ \citenamefont
  {Kalachev}(2022{\natexlab{a}})}]{Panteleev_2022}%
  \BibitemOpen
  \bibfield  {author} {\bibinfo {author} {\bibfnamefont {P.}~\bibnamefont
  {Panteleev}}\ and\ \bibinfo {author} {\bibfnamefont {G.}~\bibnamefont
  {Kalachev}},\ }\bibfield  {title} {\bibinfo {title} {Quantum {LDPC} codes
  with almost linear minimum distance},\ }\href
  {https://doi.org/10.1109/tit.2021.3119384} {\bibfield  {journal} {\bibinfo
  {journal} {{IEEE} Transactions on Information Theory}\ }\textbf {\bibinfo
  {volume} {68}},\ \bibinfo {pages} {213} (\bibinfo {year}
  {2022}{\natexlab{a}})}\BibitemShut {NoStop}%
\bibitem [{\citenamefont {Panteleev}\ and\ \citenamefont
  {Kalachev}(2022{\natexlab{b}})}]{10.1145/3519935.3520017}%
  \BibitemOpen
  \bibfield  {author} {\bibinfo {author} {\bibfnamefont {P.}~\bibnamefont
  {Panteleev}}\ and\ \bibinfo {author} {\bibfnamefont {G.}~\bibnamefont
  {Kalachev}},\ }\bibfield  {title} {\bibinfo {title} {Asymptotically good
  quantum and locally testable classical ldpc codes},\ }in\ \href
  {https://doi.org/10.1145/3519935.3520017} {\emph {\bibinfo {booktitle}
  {Proceedings of the 54th Annual ACM SIGACT Symposium on Theory of
  Computing}}},\ \bibinfo {series and number} {STOC 2022}\ (\bibinfo
  {publisher} {Association for Computing Machinery},\ \bibinfo {address} {New
  York, NY, USA},\ \bibinfo {year} {2022})\ p.\ \bibinfo {pages}
  {375–388}\BibitemShut {NoStop}%
\bibitem [{\citenamefont {{Google Quantum AI}}(2023)}]{google2023suppressing}%
  \BibitemOpen
  \bibfield  {author} {\bibinfo {author} {\bibnamefont {{Google Quantum AI}}},\
  }\bibfield  {title} {\bibinfo {title} {Suppressing quantum errors by scaling
  a surface code logical qubit},\ }\href
  {https://doi.org/10.1038/s41586-022-05434-1} {\bibfield  {journal} {\bibinfo
  {journal} {Nature}\ }\textbf {\bibinfo {volume} {614}},\ \bibinfo {pages}
  {676} (\bibinfo {year} {2023})}\BibitemShut {NoStop}%
\bibitem [{\citenamefont {Dennis}\ \emph {et~al.}(2002)\citenamefont {Dennis},
  \citenamefont {Kitaev}, \citenamefont {Landahl},\ and\ \citenamefont
  {Preskill}}]{Dennis2001}%
  \BibitemOpen
  \bibfield  {author} {\bibinfo {author} {\bibfnamefont {E.}~\bibnamefont
  {Dennis}}, \bibinfo {author} {\bibfnamefont {A.}~\bibnamefont {Kitaev}},
  \bibinfo {author} {\bibfnamefont {A.}~\bibnamefont {Landahl}},\ and\ \bibinfo
  {author} {\bibfnamefont {J.}~\bibnamefont {Preskill}},\ }\href
  {https://doi.org/10.1063/1.1499754} {\emph {\bibinfo {title} {Topological
  quantum memory}}},\ \bibinfo {type} {Tech. Rep.}\ \bibinfo {number} {9}\
  (\bibinfo {year} {2002})\BibitemShut {NoStop}%
\bibitem [{\citenamefont {Higgott}(2021)}]{higgott2021pymatching}%
  \BibitemOpen
  \bibfield  {author} {\bibinfo {author} {\bibfnamefont {O.}~\bibnamefont
  {Higgott}},\ }\href@noop {} {\bibinfo {title} {Pymatching: A python package
  for decoding quantum codes with minimum-weight perfect matching}} (\bibinfo
  {year} {2021}),\ \Eprint {https://arxiv.org/abs/2105.13082} {arXiv:2105.13082
  [quant-ph]} \BibitemShut {NoStop}%
\bibitem [{\citenamefont {Bravyi}\ \emph {et~al.}(2014)\citenamefont {Bravyi},
  \citenamefont {Suchara},\ and\ \citenamefont {Vargo}}]{Bravyi2014}%
  \BibitemOpen
  \bibfield  {author} {\bibinfo {author} {\bibfnamefont {S.}~\bibnamefont
  {Bravyi}}, \bibinfo {author} {\bibfnamefont {M.}~\bibnamefont {Suchara}},\
  and\ \bibinfo {author} {\bibfnamefont {A.}~\bibnamefont {Vargo}},\ }\bibfield
   {title} {\bibinfo {title} {Efficient algorithms for maximum likelihood
  decoding in the surface code},\ }\href
  {http://dx.doi.org/10.1103/PhysRevA.90.032326} {\bibfield  {journal}
  {\bibinfo  {journal} {Physical Review A}\ }\textbf {\bibinfo {volume} {90}},\
  \bibinfo {pages} {032326} (\bibinfo {year} {2014})}\BibitemShut {NoStop}%
\bibitem [{\citenamefont {Torlai}\ and\ \citenamefont
  {Melko}(2017)}]{PhysRevLett.119.030501}%
  \BibitemOpen
  \bibfield  {author} {\bibinfo {author} {\bibfnamefont {G.}~\bibnamefont
  {Torlai}}\ and\ \bibinfo {author} {\bibfnamefont {R.~G.}\ \bibnamefont
  {Melko}},\ }\bibfield  {title} {\bibinfo {title} {Neural decoder for
  topological codes},\ }\href {https://doi.org/10.1103/PhysRevLett.119.030501}
  {\bibfield  {journal} {\bibinfo  {journal} {Phys. Rev. Lett.}\ }\textbf
  {\bibinfo {volume} {119}},\ \bibinfo {pages} {030501} (\bibinfo {year}
  {2017})}\BibitemShut {NoStop}%
\bibitem [{\citenamefont {Varsamopoulos}\ \emph {et~al.}(2017)\citenamefont
  {Varsamopoulos}, \citenamefont {Criger},\ and\ \citenamefont
  {Bertels}}]{Varsamopoulos17}%
  \BibitemOpen
  \bibfield  {author} {\bibinfo {author} {\bibfnamefont {S.}~\bibnamefont
  {Varsamopoulos}}, \bibinfo {author} {\bibfnamefont {B.}~\bibnamefont
  {Criger}},\ and\ \bibinfo {author} {\bibfnamefont {K.}~\bibnamefont
  {Bertels}},\ }\bibfield  {title} {\bibinfo {title} {Decoding small surface
  codes with feedforward neural networks},\ }\href
  {https://doi.org/10.1088/2058-9565/aa955a} {\bibfield  {journal} {\bibinfo
  {journal} {Quantum Science and Technology}\ }\textbf {\bibinfo {volume} {3}}
  (\bibinfo {year} {2017})}\BibitemShut {NoStop}%
\bibitem [{\citenamefont {Krastanov}\ and\ \citenamefont
  {Jiang}(2017)}]{Krastanov2017}%
  \BibitemOpen
  \bibfield  {author} {\bibinfo {author} {\bibfnamefont {S.}~\bibnamefont
  {Krastanov}}\ and\ \bibinfo {author} {\bibfnamefont {L.}~\bibnamefont
  {Jiang}},\ }\bibfield  {title} {\bibinfo {title} {{Deep Neural Network
  Probabilistic Decoder for Stabilizer Codes}},\ }\href
  {https://doi.org/10.1038/s41598-017-11266-1} {\bibfield  {journal} {\bibinfo
  {journal} {Scientific Reports}\ }\textbf {\bibinfo {volume} {7}} (\bibinfo
  {year} {2017})}\BibitemShut {NoStop}%
\bibitem [{\citenamefont {Varsamopoulos}\ \emph {et~al.}(2020)\citenamefont
  {Varsamopoulos}, \citenamefont {Bertels},\ and\ \citenamefont
  {Almudever}}]{Varsamopoulos_2020}%
  \BibitemOpen
  \bibfield  {author} {\bibinfo {author} {\bibfnamefont {S.}~\bibnamefont
  {Varsamopoulos}}, \bibinfo {author} {\bibfnamefont {K.}~\bibnamefont
  {Bertels}},\ and\ \bibinfo {author} {\bibfnamefont {C.~G.}\ \bibnamefont
  {Almudever}},\ }\bibfield  {title} {\bibinfo {title} {Comparing neural
  network based decoders for the surface code},\ }\href
  {https://doi.org/10.1109/tc.2019.2948612} {\bibfield  {journal} {\bibinfo
  {journal} {{IEEE} Transactions on Computers}\ }\textbf {\bibinfo {volume}
  {69}},\ \bibinfo {pages} {300} (\bibinfo {year} {2020})}\BibitemShut
  {NoStop}%
\bibitem [{\citenamefont {Overwater}\ \emph {et~al.}(2022)\citenamefont
  {Overwater}, \citenamefont {Babaie},\ and\ \citenamefont
  {Sebastiano}}]{Overwater2022}%
  \BibitemOpen
  \bibfield  {author} {\bibinfo {author} {\bibfnamefont {R.~W.}\ \bibnamefont
  {Overwater}}, \bibinfo {author} {\bibfnamefont {M.}~\bibnamefont {Babaie}},\
  and\ \bibinfo {author} {\bibfnamefont {F.}~\bibnamefont {Sebastiano}},\
  }\bibfield  {title} {\bibinfo {title} {{Neural-Network Decoders for Quantum
  Error Correction Using Surface Codes: A Space Exploration of the Hardware
  Cost-Performance Tradeoffs}},\ }\href
  {https://doi.org/10.1109/TQE.2022.3174017} {\bibfield  {journal} {\bibinfo
  {journal} {IEEE Transactions on Quantum Engineering}\ }\textbf {\bibinfo
  {volume} {3}},\ \bibinfo {pages} {1} (\bibinfo {year} {2022})}\BibitemShut
  {NoStop}%
\bibitem [{\citenamefont {Baireuther}\ \emph {et~al.}(2018)\citenamefont
  {Baireuther}, \citenamefont {O'Brien}, \citenamefont {Tarasinski},\ and\
  \citenamefont {Beenakker}}]{Baireuther2017}%
  \BibitemOpen
  \bibfield  {author} {\bibinfo {author} {\bibfnamefont {P.}~\bibnamefont
  {Baireuther}}, \bibinfo {author} {\bibfnamefont {T.~E.}\ \bibnamefont
  {O'Brien}}, \bibinfo {author} {\bibfnamefont {B.}~\bibnamefont
  {Tarasinski}},\ and\ \bibinfo {author} {\bibfnamefont {C.~W.}\ \bibnamefont
  {Beenakker}},\ }\bibfield  {title} {\bibinfo {title}
  {Machine-learning-assisted correction of correlated qubit errors in a
  topological code},\ }\href {http://dx.doi.org/10.22331/q-2018-01-29-48}
  {\bibfield  {journal} {\bibinfo  {journal} {Quantum}\ }\textbf {\bibinfo
  {volume} {2}},\ \bibinfo {pages} {48} (\bibinfo {year} {2018})}\BibitemShut
  {NoStop}%
\bibitem [{\citenamefont {Davaasuren}\ \emph {et~al.}(2020)\citenamefont
  {Davaasuren}, \citenamefont {Suzuki}, \citenamefont {Fujii},\ and\
  \citenamefont {Koashi}}]{Davaasuren2020}%
  \BibitemOpen
  \bibfield  {author} {\bibinfo {author} {\bibfnamefont {A.}~\bibnamefont
  {Davaasuren}}, \bibinfo {author} {\bibfnamefont {Y.}~\bibnamefont {Suzuki}},
  \bibinfo {author} {\bibfnamefont {K.}~\bibnamefont {Fujii}},\ and\ \bibinfo
  {author} {\bibfnamefont {M.}~\bibnamefont {Koashi}},\ }\bibfield  {title}
  {\bibinfo {title} {{General framework for constructing fast and near-optimal
  machine-learning-based decoder of the topological stabilizer codes}},\ }\href
  {https://doi.org/10.1103/PhysRevResearch.2.033399} {\bibfield  {journal}
  {\bibinfo  {journal} {Physical Review Research}\ }\textbf {\bibinfo {volume}
  {2}} (\bibinfo {year} {2020})}\BibitemShut {NoStop}%
\bibitem [{\citenamefont {Gicev}\ \emph {et~al.}(2021)\citenamefont {Gicev},
  \citenamefont {Hollenberg},\ and\ \citenamefont {Usman}}]{Gicev2021}%
  \BibitemOpen
  \bibfield  {author} {\bibinfo {author} {\bibfnamefont {S.}~\bibnamefont
  {Gicev}}, \bibinfo {author} {\bibfnamefont {L.~C.}\ \bibnamefont
  {Hollenberg}},\ and\ \bibinfo {author} {\bibfnamefont {M.}~\bibnamefont
  {Usman}},\ }\bibfield  {title} {\bibinfo {title} {A scalable and fast
  artificial neural network syndrome decoder for surface codes},\ }\href
  {http://arxiv.org/abs/2110.05854} {\bibfield  {journal} {\bibinfo  {journal}
  {arXiv preprint arXiv:2110.05854}\ } (\bibinfo {year} {2021})}\BibitemShut
  {NoStop}%
\bibitem [{\citenamefont {Gottesman}(1997)}]{Gottesman1997}%
  \BibitemOpen
  \bibfield  {author} {\bibinfo {author} {\bibfnamefont {D.}~\bibnamefont
  {Gottesman}},\ }\href {https://arxiv.org/pdf/quant-ph/9705052.pdf} {\emph
  {\bibinfo {title} {Stabilizer codes and quantum error correction}}}\
  (\bibinfo  {publisher} {California Institute of Technology},\ \bibinfo {year}
  {1997})\BibitemShut {NoStop}%
\bibitem [{\citenamefont {Nielsen}\ and\ \citenamefont
  {Chuang}(2010)}]{nielsen_chuang_2010}%
  \BibitemOpen
  \bibfield  {author} {\bibinfo {author} {\bibfnamefont {M.~A.}\ \bibnamefont
  {Nielsen}}\ and\ \bibinfo {author} {\bibfnamefont {I.~L.}\ \bibnamefont
  {Chuang}},\ }\href {https://doi.org/10.1017/CBO9780511976667} {\emph
  {\bibinfo {title} {Quantum Computation and Quantum Information: 10th
  Anniversary Edition}}}\ (\bibinfo  {publisher} {Cambridge University Press},\
  \bibinfo {year} {2010})\BibitemShut {NoStop}%
\bibitem [{\citenamefont {Bishop}(2006)}]{Bishop}%
  \BibitemOpen
  \bibfield  {author} {\bibinfo {author} {\bibfnamefont {C.~M.}\ \bibnamefont
  {Bishop}},\ }\href@noop {} {\emph {\bibinfo {title} {Pattern Recognition and
  Machine Learning (Information Science and Statistics)}}}\ (\bibinfo
  {publisher} {Springer-Verlag},\ \bibinfo {address} {Berlin, Heidelberg},\
  \bibinfo {year} {2006})\BibitemShut {NoStop}%
\bibitem [{\citenamefont {Vaswani}\ \emph {et~al.}(2017)\citenamefont
  {Vaswani}, \citenamefont {Shazeer}, \citenamefont {Parmar}, \citenamefont
  {Uszkoreit}, \citenamefont {Jones}, \citenamefont {Gomez}, \citenamefont
  {Kaiser},\ and\ \citenamefont {Polosukhin}}]{Vaswani2017}%
  \BibitemOpen
  \bibfield  {author} {\bibinfo {author} {\bibfnamefont {A.}~\bibnamefont
  {Vaswani}}, \bibinfo {author} {\bibfnamefont {N.}~\bibnamefont {Shazeer}},
  \bibinfo {author} {\bibfnamefont {N.}~\bibnamefont {Parmar}}, \bibinfo
  {author} {\bibfnamefont {J.}~\bibnamefont {Uszkoreit}}, \bibinfo {author}
  {\bibfnamefont {L.}~\bibnamefont {Jones}}, \bibinfo {author} {\bibfnamefont
  {A.~N.}\ \bibnamefont {Gomez}}, \bibinfo {author} {\bibfnamefont
  {{\L}.}~\bibnamefont {Kaiser}},\ and\ \bibinfo {author} {\bibfnamefont
  {I.}~\bibnamefont {Polosukhin}},\ }\bibfield  {title} {\bibinfo {title}
  {Attention is all you need},\ }\href
  {https://proceedings.neurips.cc/paper/2017/file/3f5ee243547dee91fbd053c1c4a845aa-Paper.pdf}
  {\bibfield  {journal} {\bibinfo  {journal} {Advances in neural information
  processing systems}\ }\textbf {\bibinfo {volume} {30}} (\bibinfo {year}
  {2017})}\BibitemShut {NoStop}%
\bibitem [{\citenamefont {Radford}\ \emph {et~al.}(2018)\citenamefont
  {Radford}, \citenamefont {Narasimhan}, \citenamefont {Salimans},\ and\
  \citenamefont {Sutskever}}]{Radford2018ImprovingLU}%
  \BibitemOpen
  \bibfield  {author} {\bibinfo {author} {\bibfnamefont {A.}~\bibnamefont
  {Radford}}, \bibinfo {author} {\bibfnamefont {K.}~\bibnamefont {Narasimhan}},
  \bibinfo {author} {\bibfnamefont {T.}~\bibnamefont {Salimans}},\ and\
  \bibinfo {author} {\bibfnamefont {I.}~\bibnamefont {Sutskever}},\ }\bibfield
  {title} {\bibinfo {title} {Improving language understanding by generative
  pre-training}\ }(\bibinfo  {publisher} {OpenAI},\ \bibinfo {year}
  {2018})\BibitemShut {NoStop}%
\bibitem [{cha()}]{chatGPT}%
  \BibitemOpen
  \href@noop {} {}\bibinfo {note}
  {\url{https://openai.com/chatgpt}}\BibitemShut {NoStop}%
\bibitem [{\citenamefont {OpenAI}(2023)}]{openai2023gpt4}%
  \BibitemOpen
  \bibfield  {author} {\bibinfo {author} {\bibnamefont {OpenAI}},\ }\href@noop
  {} {\bibinfo {title} {Gpt-4 technical report}} (\bibinfo {year} {2023}),\
  \Eprint {https://arxiv.org/abs/2303.08774} {arXiv:2303.08774 [cs.CL]}
  \BibitemShut {NoStop}%
\bibitem [{\citenamefont {Zhang}\ and\ \citenamefont
  {Moore}(2014)}]{Zhang2014}%
  \BibitemOpen
  \bibfield  {author} {\bibinfo {author} {\bibfnamefont {P.}~\bibnamefont
  {Zhang}}\ and\ \bibinfo {author} {\bibfnamefont {C.}~\bibnamefont {Moore}},\
  }\bibfield  {title} {\bibinfo {title} {{Scalable detection of statistically
  significant communities and hierarchies, using message passing for
  modularity}},\ }\href {https://doi.org/10.1073/pnas.1409770111} {\bibfield
  {journal} {\bibinfo  {journal} {Proceedings of the National Academy of
  Sciences of the United States of America}\ }\textbf {\bibinfo {volume}
  {111}},\ \bibinfo {pages} {18144} (\bibinfo {year} {2014})}\BibitemShut
  {NoStop}%
\bibitem [{\citenamefont {Roffe}\ \emph {et~al.}(2020)\citenamefont {Roffe},
  \citenamefont {White}, \citenamefont {Burton},\ and\ \citenamefont
  {Campbell}}]{Roffe2020}%
  \BibitemOpen
  \bibfield  {author} {\bibinfo {author} {\bibfnamefont {J.}~\bibnamefont
  {Roffe}}, \bibinfo {author} {\bibfnamefont {D.~R.}\ \bibnamefont {White}},
  \bibinfo {author} {\bibfnamefont {S.}~\bibnamefont {Burton}},\ and\ \bibinfo
  {author} {\bibfnamefont {E.}~\bibnamefont {Campbell}},\ }\bibfield  {title}
  {\bibinfo {title} {Decoding across the quantum low-density parity-check code
  landscape},\ }\bibfield  {journal} {\bibinfo  {journal} {Physical Review
  Research}\ }\textbf {\bibinfo {volume} {2}},\ \href
  {https://doi.org/10.1103/PhysRevResearch.2.043423}
  {10.1103/PhysRevResearch.2.043423} (\bibinfo {year} {2020})\BibitemShut
  {NoStop}%
\bibitem [{\citenamefont {Vasmer}\ and\ \citenamefont
  {Browne}(2019)}]{PhysRevA.100.012312}%
  \BibitemOpen
  \bibfield  {author} {\bibinfo {author} {\bibfnamefont {M.}~\bibnamefont
  {Vasmer}}\ and\ \bibinfo {author} {\bibfnamefont {D.~E.}\ \bibnamefont
  {Browne}},\ }\bibfield  {title} {\bibinfo {title} {Three-dimensional surface
  codes: Transversal gates and fault-tolerant architectures},\ }\href
  {https://doi.org/10.1103/PhysRevA.100.012312} {\bibfield  {journal} {\bibinfo
   {journal} {Phys. Rev. A}\ }\textbf {\bibinfo {volume} {100}},\ \bibinfo
  {pages} {012312} (\bibinfo {year} {2019})}\BibitemShut {NoStop}%
\bibitem [{git()}]{github}%
  \BibitemOpen
  \href@noop {} {}\bibinfo {note}
  {\url{https:github.com/CHY-i/qecGPT}}\BibitemShut {NoStop}%
\bibitem [{\citenamefont {Kolmogorov}(2009)}]{Kolmogorov2009}%
  \BibitemOpen
  \bibfield  {author} {\bibinfo {author} {\bibfnamefont {V.}~\bibnamefont
  {Kolmogorov}},\ }\bibfield  {title} {\bibinfo {title} {{Blossom V: A new
  implementation of a minimum cost perfect matching algorithm}},\ }\href
  {https://doi.org/10.1007/s12532-009-0002-8} {\bibfield  {journal} {\bibinfo
  {journal} {Mathematical Programming Computation}\ }\textbf {\bibinfo {volume}
  {1}},\ \bibinfo {pages} {43} (\bibinfo {year} {2009})}\BibitemShut {NoStop}%
\bibitem [{\citenamefont {Fakoor}\ \emph {et~al.}(2020)\citenamefont {Fakoor},
  \citenamefont {Chaudhari}, \citenamefont {Mueller},\ and\ \citenamefont
  {Smola}}]{fakoor2020trade}%
  \BibitemOpen
  \bibfield  {author} {\bibinfo {author} {\bibfnamefont {R.}~\bibnamefont
  {Fakoor}}, \bibinfo {author} {\bibfnamefont {P.}~\bibnamefont {Chaudhari}},
  \bibinfo {author} {\bibfnamefont {J.}~\bibnamefont {Mueller}},\ and\ \bibinfo
  {author} {\bibfnamefont {A.~J.}\ \bibnamefont {Smola}},\ }\href@noop {}
  {\bibinfo {title} {Trade: Transformers for density estimation}} (\bibinfo
  {year} {2020}),\ \Eprint {https://arxiv.org/abs/2004.02441} {arXiv:2004.02441
  [cs.LG]} \BibitemShut {NoStop}%
\bibitem [{\citenamefont {Williams}(1992)}]{williams1992simple}%
  \BibitemOpen
  \bibfield  {author} {\bibinfo {author} {\bibfnamefont {R.~J.}\ \bibnamefont
  {Williams}},\ }\bibfield  {title} {\bibinfo {title} {Simple statistical
  gradient-following algorithms for connectionist reinforcement learning},\
  }\href@noop {} {\bibfield  {journal} {\bibinfo  {journal} {Machine learning}\
  }\textbf {\bibinfo {volume} {8}},\ \bibinfo {pages} {229} (\bibinfo {year}
  {1992})}\BibitemShut {NoStop}%
\bibitem [{\citenamefont {Wu}\ \emph {et~al.}(2019)\citenamefont {Wu},
  \citenamefont {Wang},\ and\ \citenamefont {Zhang}}]{PhysRevLett.122.080602}%
  \BibitemOpen
  \bibfield  {author} {\bibinfo {author} {\bibfnamefont {D.}~\bibnamefont
  {Wu}}, \bibinfo {author} {\bibfnamefont {L.}~\bibnamefont {Wang}},\ and\
  \bibinfo {author} {\bibfnamefont {P.}~\bibnamefont {Zhang}},\ }\bibfield
  {title} {\bibinfo {title} {Solving statistical mechanics using variational
  autoregressive networks},\ }\href
  {https://doi.org/10.1103/PhysRevLett.122.080602} {\bibfield  {journal}
  {\bibinfo  {journal} {Phys. Rev. Lett.}\ }\textbf {\bibinfo {volume} {122}},\
  \bibinfo {pages} {080602} (\bibinfo {year} {2019})}\BibitemShut {NoStop}%
\bibitem [{\citenamefont {Bombin}\ \emph {et~al.}(2012)\citenamefont {Bombin},
  \citenamefont {Andrist}, \citenamefont {Ohzeki}, \citenamefont {Katzgraber},\
  and\ \citenamefont {Martin-Delgado}}]{Bombin_2012}%
  \BibitemOpen
  \bibfield  {author} {\bibinfo {author} {\bibfnamefont {H.}~\bibnamefont
  {Bombin}}, \bibinfo {author} {\bibfnamefont {R.~S.}\ \bibnamefont {Andrist}},
  \bibinfo {author} {\bibfnamefont {M.}~\bibnamefont {Ohzeki}}, \bibinfo
  {author} {\bibfnamefont {H.~G.}\ \bibnamefont {Katzgraber}},\ and\ \bibinfo
  {author} {\bibfnamefont {M.~A.}\ \bibnamefont {Martin-Delgado}},\ }\bibfield
  {title} {\bibinfo {title} {Strong resilience of topological codes to
  depolarization},\ }\bibfield  {journal} {\bibinfo  {journal} {Physical Review
  X}\ }\textbf {\bibinfo {volume} {2}},\ \href
  {https://doi.org/10.1103/physrevx.2.021004} {10.1103/physrevx.2.021004}
  (\bibinfo {year} {2012})\BibitemShut {NoStop}%
\bibitem [{\citenamefont {Battistel}\ \emph {et~al.}(2023)\citenamefont
  {Battistel}, \citenamefont {Chamberland}, \citenamefont {Johar},
  \citenamefont {Overwater}, \citenamefont {Sebastiano}, \citenamefont
  {Skoric}, \citenamefont {Ueno},\ and\ \citenamefont
  {Usman}}]{Battistel2023RealTimeDF}%
  \BibitemOpen
  \bibfield  {author} {\bibinfo {author} {\bibfnamefont {F.}~\bibnamefont
  {Battistel}}, \bibinfo {author} {\bibfnamefont {C.}~\bibnamefont
  {Chamberland}}, \bibinfo {author} {\bibfnamefont {K.}~\bibnamefont {Johar}},
  \bibinfo {author} {\bibfnamefont {R.~W.}\ \bibnamefont {Overwater}}, \bibinfo
  {author} {\bibfnamefont {F.}~\bibnamefont {Sebastiano}}, \bibinfo {author}
  {\bibfnamefont {L.}~\bibnamefont {Skoric}}, \bibinfo {author} {\bibfnamefont
  {Y.}~\bibnamefont {Ueno}},\ and\ \bibinfo {author} {\bibfnamefont
  {M.}~\bibnamefont {Usman}},\ }\bibfield  {title} {\bibinfo {title} {Real-time
  decoding for fault-tolerant quantum computing: Progress, challenges and
  outlook},\ }\href {https://arxiv.org/pdf/2303.00054.pdf} {\bibfield
  {journal} {\bibinfo  {journal} {arXiv preprint arXiv:2303.00054}\ } (\bibinfo
  {year} {2023})}\BibitemShut {NoStop}%
\bibitem [{\citenamefont {Bombin}\ and\ \citenamefont
  {Martin-Delgado}(2007)}]{Bombin2007}%
  \BibitemOpen
  \bibfield  {author} {\bibinfo {author} {\bibfnamefont {H.}~\bibnamefont
  {Bombin}}\ and\ \bibinfo {author} {\bibfnamefont {M.~A.}\ \bibnamefont
  {Martin-Delgado}},\ }\bibfield  {title} {\bibinfo {title} {Optimal resources
  for topological two-dimensional stabilizer codes: Comparative study},\ }\href
  {https://doi.org/10.1103/PhysRevA.76.012305} {\bibfield  {journal} {\bibinfo
  {journal} {Physical Review A}\ }\textbf {\bibinfo {volume} {76}} (\bibinfo
  {year} {2007})}\BibitemShut {NoStop}%
\end{thebibliography}
\end{document}